\documentclass[a4paper]{elsarticle}
\usepackage[margin=0.75in]{geometry}
\usepackage{lineno,hyperref}
\usepackage{upgreek} % non italicised greek letters
\usepackage{amsmath} % for bmatrix among others
\usepackage{amssymb}
\usepackage{dirtytalk} % for quotation 
\usepackage{tabularx}
\usepackage{multirow} % for spanning rows
\usepackage{cleveref} % for cref
\usepackage{booktabs} % top rule table
\usepackage{float}
\usepackage{longtable}
\usepackage[title]{appendix}
\usepackage{soul} % for \hl
\usepackage{framed}
\usepackage{multicol}

\immediate\write18{%
makeindex -s nomencl.ist -o \jobname.nls -t \jobname.nlg \jobname.nlo%
}
\usepackage{nomencl}
\makenomenclature
\DeclareMathOperator*{\argmax}{arg\,max}

\graphicspath{{/Users/floracharbonnier/OneDrive - Nexus365/DPhil/Writing}{/Users/floracharbonnier/OneDrive - Nexus365/DPhil/Writing/RLPaper}}

\modulolinenumbers[5]

\journal{Applied Energy}

\usepackage{etoolbox}
\AtBeginEnvironment{appendices}{\crefalias{section}{appendix}}

\begin{document}

\begin{frontmatter}

\title{Scalable multi-agent reinforcement learning for distributed control of residential energy flexibility}

%% authors
\author[affil1]{Flora Charbonnier\corref{mycorrespondingauthor}}
\cortext[mycorrespondingauthor]{Corresponding author}
\ead{flora.charbonnier@eng.ox.ac.uk}

\author[affil2]{Thomas Morstyn}

\author[affil1]{Malcolm D. McCulloch}

\address[affil1]{Department of Engineering Science, University of Oxford, UK}
\address[affil2]{School of Engineering, University of Edinburgh, UK}

\begin{abstract}
This paper proposes a novel scalable type of multi-agent reinforcement learning-based coordination for distributed residential energy. Cooperating agents learn to control the flexibility offered by electric vehicles, space heating and flexible loads in a partially observable stochastic environment. In the standard independent Q-learning approach, the coordination performance of agents under partial observability drops at scale in stochastic environments. Here, the novel combination of learning from off-line convex optimisations on historical data and isolating marginal contributions to total rewards in reward signals increases stability and performance at scale.  Using fixed-size Q-tables, prosumers are able to assess their marginal impact on total system objectives without sharing personal data either with each other or with a central coordinator. Case studies are used to assess the fitness of different combinations of exploration sources, reward definitions, and multi-agent learning frameworks.  It is demonstrated that the proposed strategies create value at individual and system levels thanks to reductions in the costs of energy imports, losses, distribution network congestion, battery depreciation and greenhouse gas emissions.
\end{abstract}

\end{frontmatter}

\section*{Highlights}
\begin{itemize}
\item Privacy-preserving multi-agent reinforcement learning is used to coordinate residential energy  
\item Learning from optimisations improves coordination scalability in stochastic environments
\item Marginal reward signals further enhance cooperation relative to previous approaches
\item The curse of dimensionality is mitigated by the use of fixed-size Q-tables
\item Case studies with large real-life datasets yield 33.7\% local and global cost reductions 
\end{itemize}

\begin{multicols}{2}
%\linenumbers

\section{Introduction}
This paper addresses the scalability issue of distributed domestic energy flexibility coordination in a cost-efficient and privacy-preserving manner.  A novel class of coordination strategies using optimisation-based multi-agent reinforcement learning (MARL\footnote{A full nomenclature is available in \Cref{app:nomenclature}}) with fixed Q-table size is proposed for household-level decision-making, tackling the challenge of scalability for simultaneously learning independent agents under partial observability in a stochastic environment \citep{Matignon2012}. Multiple versions of the novel strategy are assessed to maximise the statistical expectation of system-wide benefits, including local battery costs, grid costs and greenhouse gas emissions. 

Widespread electrification of primary energy provision and decarbonisation of the power sector are two vital prerequisites for limiting anthropogenic global warming to 1.5$^o$C above pre-industrial levels. To reduce risks of climate-related impacts on health, livelihood, security and economic growth, intermittent renewable power supplies could be required to supply 70\% to 85\% of electricity by 2050 \citep{IPCC2015}. However, this poses the challenges of the intermittency and limited controllability of resources \citep{Bose2019}.  Therefore, a robust, decarbonised power system will rely on two structural features: decentralisation and demand response (DR) \citep{Leautier2019}. The coordination of distributed flexible energy resources can help reduce costs for transmission, storage,  peaking plants and capacity reserves,  improve grid stability, align demand with decarbonised energy provision, promote energy independence and security, and lower household energy bills \citep{Vazquez-Canteli2019, Pumphrey2020}.

Residential sites constitute a significant share of potential DR, representing for example 38.5\% of the 2019 UK electricity demand, and 56.4\% of energy consumption if including transport and heat, which are both undergoing electrification \citep{BEIS2021}. Increasing ownership of EVs and PV panels has been facilitated by regulatory changes, with many countries committing to internal combustion car phase-outs in the near future, and by plummeting costs, with an 82\% and 87\% levelised cost drop between 2010 and 2019 for EVs and PV panels \citep{Agency2018,BloomberNEF2019}. This potential is so far underexploited, as DR primarily focuses on larger well-known industrial and commercial actors that require less coordination and data management \citep{CharlesRiverAssociates2017}, with most customers still limited to trade with utility companies \citep{Chen2019}. The primary hurdles to unlocking residential flexibility are the high capital cost of communication and control infrastructure as the domestic potential is highly fragmented \citep{Leautier2019}, concerns about privacy and hindrance of activities \citep{Bugden2019,Pumphrey2020}, and computational challenges for real-time control at scale \citep{Moret2019}.  

Traditionally, convex optimisation would be used to maximise global coordination objectives in convex problems with variables known ahead of time. Techniques such as least-squares and linear programming have been well-studied for over a century \citep{Boyd2009}. However, residential energy coordination presents challenges to its application. Firstly,  optimisations that are centralised are hindered by privacy, acceptance, and communication constraints, and present exponential time complexity at the scale of millions of homes \citep{Dasgupta2016}. Secondly, standard optimisation methods cannot be used without full knowledge of the system's inputs and dynamics \citep{Recht2018}. In residential energy, agents only have partial observability of the system due to both the stochasticity and uncertainty of environment variables such as individual residential consumption and generation profiles, and to the privacy and infrastructure cost constraints that hinder communication between agents during implementation \citep{FrancoisLavet2017}. Not relying on shared information may also improve the robustness of the solutions to failure of other agents, communication delays, and unreliable information, and improve adaptability to changing environments \citep{Sen1994}. Finally, the real-life complex electricity grid environment may not be amenable to a convex model representation. Due to the heterogeneity of users and behaviours needing different parameters and models,  the large-scale use of model-based controllers is cumbersome \citep{Ruelens2017}. A model-free approach instead avoids modelling non-trivial interactions of parameters, including private information \citep{Dasgupta2016}.  

Given these challenges to residential energy flexibility coordination, and the specific constraints of the problem at play which renders traditional approaches unsuitable, we seek to develop a novel coordination mechanism which satisfies the following criteria,  as tested in real-life scenarios:
\begin{itemize}
\item Computational scalability: minimal and constant computation burden during implementation as the system size increases; 
\item Performance scalability: no drop in coordination performance as the system size increases, measured in savings obtained per hour and per agent;
\item Acceptability: local control of appliances, no communication of personal data, thermal discomfort, or hindrance/delay of activities.
\end{itemize}

The rest of this paper is organised as follows. In \Cref{sec:gapanalysis} we motivate the novel MARL approach with a literature review and a gap analysis. In \Cref{system}, a system model is presented that includes household-level modelling of EVs, space heating, flexible loads and PV generation.  \Cref{RLSection} lays out the MARL methodology, with various methodological options for independent agents to learn to cooperate. In \Cref{data}, the input data used to populate the model is presented. In \Cref{results}, the performance of different MARL strategies is compared to lower and upper bounds in case studies. Finally, we conclude in \Cref{conclusion}.

\section{MARL-based energy coordination: literature review and gap analysis}\label{sec:gapanalysis}
Reinforcement learning (RL) can overcome the constraints faced by centralised convex optimisation for residential energy coordination, by allowing for decentralised and model-free decision-making based on partial knowledge. RL is an artificial intelligence (AI) framework for goal-oriented agents\footnote{Here agents are independent computer systems acting on behalf of prosumers \citep{Wooldridge2002}. Prosumers are proactive consumers with distributed energy resources actively managing their consumption, production and storage of energy \citep{Morstyn2018_Federated}. } to learn sequential decision-making by interacting with an uncertain environment \citep{Sutton1998}.  As an increasing wealth of data is collected in local electricity systems, RL is of growing interest for the real-time coordination of distributed energy resources (DERs) \citep{Antonopoulos2020,Vazquez-Canteli2019}.  Instead of optimising based on inherently uncertain data, RL more realistically searches for statistically optimal sequential decisions given partial observation and uncertainty, with no \emph{a priori} knowledge \citep{Recht2018}.  Approximate learning methods may be more computationally scalable, more efficient in exploring high-dimensional state spaces and therefore more scalable than exact global optimisation with exponential time complexity \citep{Schellenberg2020, Dasgupta2016}. 

As classified in \citep{CharbonnierReview}, numerous RL-based coordination methods have been proposed in the literature for residential energy coordination, though with remaining limitations in terms of scalability and privacy protection. On the one hand, in RL-based direct control strategies, a central controller directly controls individual units, and households directly forfeit their data and control to a central RL-based scheduler \citep{ONeill2010}. While most existing AI-based DR research thus assumes fully observable tasks \citep{Antonopoulos2020}, direct controllability of resources from different owners with different objectives and resources and subject to privacy, comfort and security concerns is challenging \citep{Darby2020}. Moreover, centralised policies do not scale due to the curse of dimensionality as the state and action spaces grow exponentially with the system size \citep{Powell2011}.  On the other hand, RL-based indirect control strategies consider decision-making at the prosumer level, entering the realm of MARL. This can be achieved using different communication structures, with either centralised, bilateral, or no sharing of personal information, as presented below.

Firstly, agents may share information with a central entity,  which in turn broadcasts signals based on a complete picture of the coordination problem. For example, the central entity may send unidirectional price signals to customers based on information such as prosumers’ costs, constraints and day-ahead forecasts. RL can inform both the dynamic price signal \citep{Lu2019, Kim2016}, and the prosumer response to price signals \citep{Kim2016,Babar2018}. The central entity may also collect competitive bids and set trades and match prosumers centrally, where RL algorithms are used to refine individual bidding strategies \citep{Vaya2014, Ye2020, Dauer2013,Sun2015,Kim2020} or to dictate the auction market clearing \citep{Chen2019,Claessens2013}. Units may also use RL to cooperate towards common objectives with the mediation of a central entity that redistributes centralised personal information \citep{Zhang2017,Dusparic2015,Dusparic2013,Hurtado2018}. However, information centralisation also raises costs, security, privacy and scalability of computation issues. Biased information may lead to inefficient or even infeasible decisions \citep{Morstyn2020_P2P}. 

Secondly, RL-based coordination has been proposed where prosumers only communicate information bilaterally without a central authority. For example, in \citep{Taylor2014} agents use transfer learning with distributed W-learning to achieve local and system objectives. Bilateral peer-to-peer communication offers autonomy and expression of individual preferences, though with remaining risks around privacy and bounded rationality \citep{Herbert1982}. There is greater robustness to communication failures compared situations with a single point of failure. However, as the system size increases,  the number of communication iterations until algorithmic convergence increases, requiring adequate computational resources and limited communication network latency for feasibility \citep{Guerrero2020}. The safe way of implementing distributed transactions to ensure data protection is an ongoing subject of research \citep{CharbonnierReview}. 

Finally,  in RL-based implicit coordination strategies, prosumers rely solely on local information to make decisions. For example, in \citep{Cao2019, Yang2019}, competitive agents in isolation maximise their profits in RL-based energy arbitrage, though they do not consider the impacts of individual actions on the rest of the system, with potential negative impacts for the grid. For example, a concern is that all loads receive the same incentive, the natural diversity on which the grid relies may be diminished \citep{Crozier2018_Mitigating}, and the peak potentially merely displaced, with overloads on upstream transformers. Implicit cooperation, which keeps personal information at the local level while encouraging cooperation towards global objectives, has been thus far under-researched beyond frequency control.  In \citep{Rozada2020}, agents learn the optimal way of acting and interacting with the environment to restore frequency using local information only. This is a promising approach for decentralised control. However, the applicability in more complex scenarios with residential electric vehicles and smart heating load scheduling problems has not been considered. Moreover, the convergence slows down for increasing number of agents, and scalability beyond 8 agents has not been investigated. Indeed, fundamental challenges to the coordination of simultaneously learning independent agents at scale under partial observability in a stochastic environment have been identified when using traditional RL algorithms [1]: independent learners may reach individual policy equilibriums that are incompatible with a global Pareto optimal, the non-stationarity of the environment due to other concurrently learning agents affects convergence, and the stochasticity of the environment prevents agents from discriminating between their own contribution to global rewards and noise from other agents or the environment. Novel methods are therefore needed to develop this approach.

We seek to bridge this gap, using implicit coordination to unlock the so-far largely untapped value from residential energy flexibility to provide both individual and system benefits. We propose a new class of MARL-based implicit cooperation strategies for residential DR, to make the best use of the flexibility offered by increasingly accessible assets such as photovoltaic (PV) panels, electric vehicle (EV) batteries, smart heating and flexible loads. Agents learn RL policies using a data-based, model-free statistical approach by exploring a shared environment and interacting with decentralised partially observable Markov decision processes (Dec-POMDPs), either through random exploration or learning from convex optimisation results. In the first rehearsal phase \citep{Kraemer2016} with full understanding of the system, they learn to cooperate to reach system-wide benefits by assessing the global impact of their individual actions, searching for trade-offs between local, grid and social objectives. The pre-learned policies are then used to make decisions under uncertainty given limited local information only.

This approach satifies the computational scalability, coordination scalability and acceptance criteria set out in this paper.

Firstly,  the real-time control method is computationally scalable thanks to fixed-size Q-tables which avoid the curse of dimensionality, and there is only minimal, constant local computation required to implement the pre-learned policies during implementation. No further communication is required for implementation. This increases robustness to communication issues and data inaccuracy relative to when relying on centralised and bilateral communication, and cuts the costs of household computation and two-way communication infrastructure. 

Secondly, we address the outstanding MARL coordination performance scalability issue for agents with partial observability in a stochastic environment seeking to maximise rewards which also depend on other concurrently learning agents \citep{Busoniu2008,Matignon2012}. The case studies in this paper show that allowing agents to learn from omniscient, stable, and consistent optimisation solutions can successfully act as an equilibrium-selection mechanism, while the use of marginal rewards improves learnability\footnote{\say{the sensitivity of an agent’s utility to its own actions as opposed to actions of others, which is often low in fully cooperative Markov games} \citep{Matignon2012}} by isolating individual contributions to global rewards. This novel methodological combination offers significant improvements on MARL scalability and convergence issues, with high coordination performance maintained as the number of agents increases, where that of standard MARL drops at scale. 

Finally,  this method tackles acceptability issues, with no interference in personal comfort nor communication of personal data. 

The specific novel contributions of this paper are (a) a novel class of decentralised flexibility coordination strategies, MARL-based implicit cooperation, with no communication and fixed-size Q-tables to mitigate the curse of dimensionality; (b) a novel MARL exploration strategy for agents under partial observability to learn from omniscient, convex optimisations prior to implementation for convergence to robust cooperation at scale; and (c) the design and testing with large banks of real-world data of combinations of reward definitions, exploration strategies and multi-agent learning frameworks for assessing individual impacts on global energy, grid and storage costs. Methodologies are identified which outperform a baseline with increasing numbers of agents despite uncertainty.

\section{Local system description}\label{system}
% this figure is from the 21.01.19 supervisor meeting slides
\begin{figure*}[!t]
\begin{center}
\includegraphics[width=0.7\linewidth]{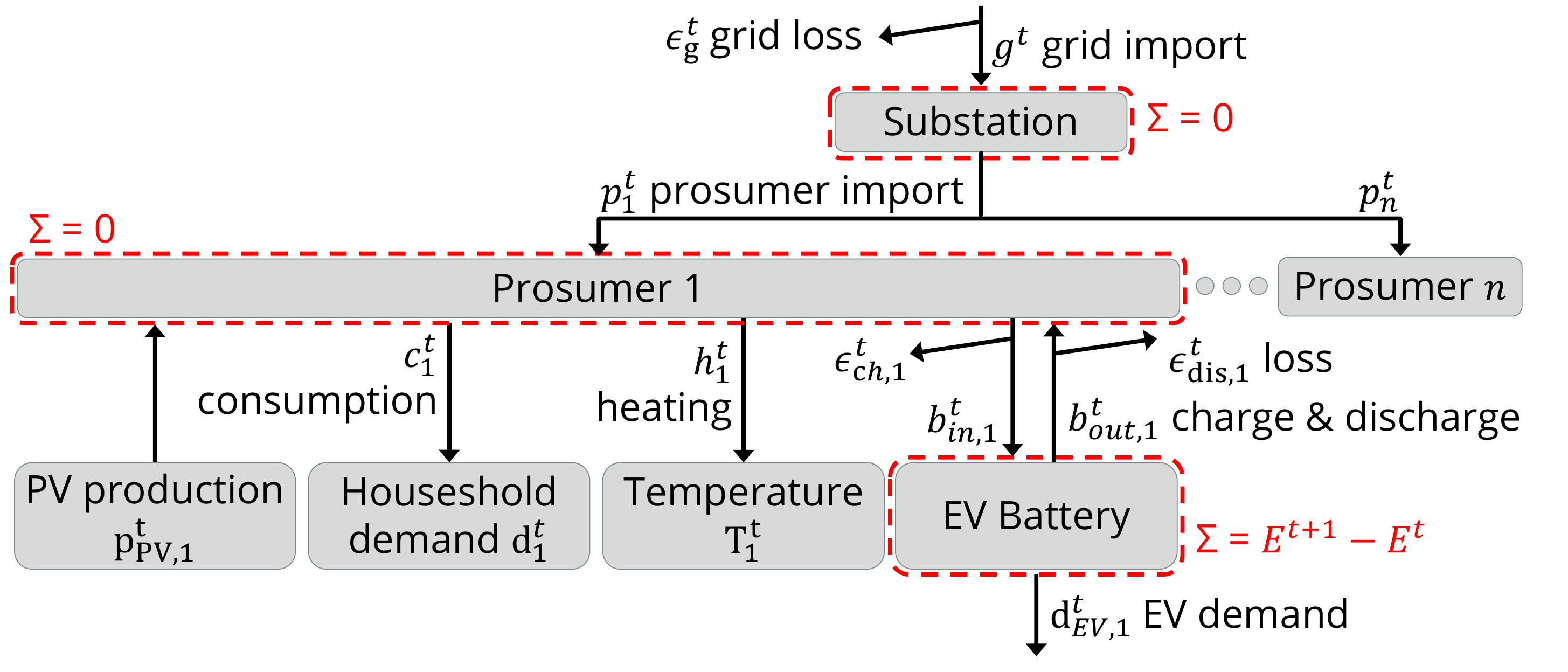}

\end{center}
\caption{Local system model. Red dotted lines denote energy balances.}
\label{fig:EnergyBalance}
\end{figure*}

In this section, the variables, objective function and constraints of the problem are described. This sets the frame for the application of the RL algorithms presented in \Cref{RLSection}.

\subsection{Variables}\label{variables}
We consider a set of time steps $t \in \mathcal{T} = \{t_0,...,t_\textrm{end}\}$ and a set of prosumers $i \in \mathcal{P} = \{1,...,n\}$. Decision variables are \emph{italicised} and input data are written in roman. Energy units are used unless specified otherwise. Participants have an EV, a PV panel, electric space heating and generic flexible loads.

The EV at-home availability $\upmu_i^t$ (1 if available, 0 otherwise), EV demand for required trips $\textrm{d}_{\textrm{EV},i}^t$, household electric demand $\textrm{d}_i^{t}$,  PV production $\textrm{p}_{\textrm{PV},i}^t$, external temperature $\textrm{T}_{\textrm{e}}^t$ and solar heat flow rate $\upphi^t$ are specified as inputs for $t \in \mathcal{T}$ and $i \in \mathcal{P}$.

The local decisions by prosumers are the energy flows in and out of the battery $b_{\textrm{in},i}^t$ and $b_{\textrm{out},i}^t$, the electric heating consumption $h_i^t$ and the prosumer consumption $c_i^t$. These have both local and system impacts (\Cref{fig:EnergyBalance}). Local impacts include battery energy levels $E_i^t$,  losses $\epsilon_{\textrm{ch},i}^t$ and $\epsilon_{\textrm{dis},i}^t$,  prosumer import $p_i^t$, building mass temperature $T_{\textrm{m},i}^t$ and indoor air temperature $T_{\textrm{air},i}^t$. System impacts arise through the costs of total grid import $g^t$ and distribution network trading. Distribution network losses and reactive power flows are not included.  

\subsection{Objective function}\label{objfunc}

Prosumers cooperate to minimise system costs consisting of grid ($c_\textrm{g}^t$), distribution ($c_\textrm{d}^t$) and storage ($c_\textrm{s}^t$) costs. This objective function will be maximised both in convex optimisations off-line -- to provide an upper bound for the achievable objective function, and in some cases to provide information to the learners during the simulated learning phase -- and in the learning of MARL policies for decentralised online implementation. 

\begin{equation}
\max F = \sum_{\forall t \in \mathcal{T}}{\hat{F}_t} = \sum_{\forall t \in \mathcal{T}}{- (c_\textrm{g}^t + c_\textrm{d}^t + c_\textrm{s}^t )}
\end{equation}

\begin{equation}
c_\textrm{g}^t = \textrm{C}_\textrm{g}^t \left( g^t + \epsilon_g \right)
\end{equation}
Where losses incurred by imports and exports from and to the main grid are approximated as
\begin{equation}
\epsilon_g = \frac{\textrm{R}}{\textrm{V}^2}\left(g^t\right)^2
\end{equation}

The grid cost coefficient $\textrm{C}_\textrm{g}^t$ is the sum of the grid electricity price and the product of the carbon intensity of the generation mix at time $t$ and the Social Cost of Carbon which reflects the long-term societal cost of emitting greenhouse gases \citep{ParryM}. The impacts of local decisions on upstream energy prices are neglected.  Grid losses are approximated using the nominal root mean square grid voltage $\textrm{V}$ and the average resistance between the main grid and the distribution network $\textrm{R}$ \citep{Multiclass}, based on the assumption of small network voltage drops and relatively low reactive power flows \citep{Coffrin2012}. The second-order dependency disincentivises large power imports and exports, which helps ensure interactions of transmission and distribution networks do not reduce system stability.

\begin{equation}
c_\textrm{d}^t = \textrm{C}_\textrm{d}\sum_{i \in \mathcal{P}}{\max\left(- p_i^t,0\right)}
\end{equation}
\noindent Distribution costs $c_\textrm{d}^t$ are proportional to the distribution charge $\textrm{C}_\textrm{d}$ on exports. The resulting price spread between individual imports and exports decreases risks of network constraints violation by incentivising the use of local flexibility first \citep{Morstyn2020_IntegratingP2P}. Distribution network losses due to power flows between prosumers are neglected so there is no second-order dependency.
\begin{equation}
c_\textrm{s}^t = \textrm{C}_\textrm{s}\sum_{i \in \mathcal{P}}{\left(b_{\textrm{in},i}^t + b_{\textrm{out},i}^t\right)}
\end{equation}

\noindent Storage battery depreciation costs $c_\textrm{s}^t$ are assumed to be proportional to throughput using the depreciation coefficient $\textrm{C}_\textrm{s}$, assuming a uniform energy throughput degradation rate \citep{DufoLopez2014}.

\subsection{Constraints}

Let $\textrm{E}_0$,  $\underline{\textrm{E}}$ and $\overline{\textrm{E}}$ be the initial, minimum and maximum battery energy levels, $\upeta_\textrm{ch}$ and $\upeta_\textrm{dis}$ the charge and discharge efficiencies, and $\overline{\textrm{b}_\textrm{in}}$ the maximum charge per time step. Demand $\textrm{d}_{i,k}^{t_\textrm{D}}$ is met by the sum of loads consumed $\hat{c}_{i,k,t_\textrm{C},t_\textrm{D}}$ at time $t_\textrm{C}$ by prosumer $i$ for load of type $k$ (fixed or flexible) demanded at $t_\textrm{D}$. The flexibility boolean $\textrm{f}_{i,k,t_\textrm{C},t_\textrm{D}}$ indicates if time $t_\textrm{C}$ lies within the acceptable range to meet $\textrm{d}_{i,k}^{t_\textrm{D}}$. A Crank-Nicholson scheme \citep{ISO2007} is employed to model heating, with $\upkappa$ a 2x5 matrix of temperature coefficients, and $\underline{\textrm{T}}_i^t$ and $\overline{\textrm{T}}_i^t$ lower and upper temperature bounds. System constraints for steps $\forall \ t \in \mathcal{T}$ and prosumers $\forall \ i \in \mathcal{P}$ are:

\begin{itemize}
\item Prosumer and substation energy balance (see \Cref{fig:EnergyBalance})
\begin{equation}
	p_i^t = c_i^t + h_i^t + \frac{b_{\textrm{in},i}^t}{\upeta_\textrm{ch}} - {\upeta_\textrm{dis}} b_{\textrm{out},i}^t - \textrm{p}_{\textrm{PV},i}^t 
\end{equation}
\begin{equation}
\sum_{i \in \mathcal{P}}{p_i^t} = g^t
\end{equation}
\item Battery energy balance 
\begin{equation}
	E_i^{t+1} = E_i^t  + b_{\textrm{in},i}^t - b_{\textrm{out},i}^t - \textrm{d}_{\textrm{EV},i}^t 
\end{equation}
\item Battery charge and discharge constraints
\begin{equation}
	 \textrm{E}_0 = E_i^{t_0}  = E_i^{t_\textrm{end}} + b_{\textrm{in},i}^{t_{\textrm{end}}} - b_{\textrm{out},i}^{t_{\textrm{end}}} - \textrm{d}_{\textrm{EV},i}^{t_{\textrm{end}}} 
\end{equation}
\begin{equation}
	\upmu_i^t\underline{\textrm{E}}_i \leq E_i^t \leq \overline{\textrm{E}}_i
\end{equation}
\begin{equation}
	b_{\textrm{in},i}^t \leq \upmu_i^t \overline{\textrm{b}_\textrm{in}}
\end{equation}
\begin{equation}
	b_{\textrm{out},i}^t \leq \upmu_i^t \overline{\textrm{E}}_i
\end{equation}

\item Consumption flexibility --- the demand of type $k$ at time $t_\textrm{D}$ by prosumer $i$ must be met by the sum of partial consumptions $\hat{c}_{i,k,t_\textrm{C},t_\textrm{D}}$ at times $t_\textrm{C}...t_\textrm{C}+\textrm{n}_\textrm{flex}$ within the time frame $\textrm{n}_\textrm{flex}$ specified by the flexibility of each type of demand in matrix $\textrm{f}_{i,k,t_\textrm{C},t_\textrm{D}}$
\begin{equation}\label{eq:demandmet}
	\sum_{t_\textrm{C}\in\mathcal{T}}{\hat{c}_{i,k,t_\textrm{C},t_\textrm{D}} \textrm{f}_{i,k,t_\textrm{C},t_\textrm{D}}} = \textrm{d}_{i,k}^{t_\textrm{D}} 
\end{equation}
\item Consumption --- the total consumption at time $t_\textrm{C}$ is the sum of all partial consumptions $\hat{c}_{i,k,t_\textrm{C},t_\textrm{D}}$ meeting parts of demands from current and previous time steps $t_\textrm{D}$:
\begin{equation}\label{eq:totalcons}
	\sum_{t_\textrm{D}\in\mathcal{N}}{\hat{c}_{i,k,t_\textrm{C},t_\textrm{D}}}= c_{i,k}^{t_\textrm{C}} 
\end{equation}

\item Heating --- the workings to obtain this equation are included in \Cref{app:heating}:
\begin{equation}\label{eq:main_heating}
\begin{bmatrix}
T_{\textrm{m},i}^{t+1}\\
T_{\textrm{air},i}^{t+1} 
\end{bmatrix}
 = \upkappa 
 \begin{bmatrix}
1,
T_{\textrm{m},i}^{t},
\textrm{T}_{\textrm{e}}^t,
\upphi^t,
h_i^t
\end{bmatrix}^\intercal
\end{equation}
\begin{equation}
\underline{\textrm{T}}_i^t \leq T_{\textrm{air},i}^t \leq \overline{\textrm{T}}_i^t
\end{equation}
\item Non-negativity constraints 
\begin{equation}
	c_i^t, h_i^t,E_i^t, b_{\textrm{in},i}^t, b_{\textrm{out},i}^t, \hat{c}_{i,l,t_\textrm{C},t_\textrm{D}} \geq 0
\end{equation}
\end{itemize}

While the proposed framework could accommodate the use of idiosyncratic satisfaction functions to perform trade-offs between flexibility use and users' comfort, no such trade-offs are considered in this paper, with comfort requirements for temperature and EV usage always being met. Field evaluations have shown that programmes that do not maintain thermal comfort are consistently overridden, increasing overall energy use and costs \citep{Sachs2012}, while interference in consumption patterns and temperature set-points cause dissatisfaction \citep{Vazquez-Canteli2019}.  Meeting fixed domestic loads,  ensuring sufficient charge for EV trips, and maintaining comfortable temperatures are therefore set constraints. 

\section{Reinforcement learning methodology}\label{RLSection}
The MARL approach is now presented in which independent prosumers learn to make individual decisions which together maximise the statistical expectation of the objective function in \Cref{system}.  

At time step $t \in \mathcal{T}$, each agent is in a state $s_i^t \in \mathcal{S}$ corresponding to accessible observations (here the time-varying grid cost), and selects an action $a_i^t \in \mathcal{A}$ as defined in \Cref{sec:agentdecision}. This action dictates the decision variables in \Cref{variables} $b_{\textrm{in},i}^t$, $b_{\textrm{out},i}^t$, $h_i^t$ and $c_i^t$. The environment then produces a reward $r^t \in \mathcal{R}$ which corresponds to the share $\hat{F}_t$ of the system objective function presented in \Cref{objfunc} and agents transition to a state $s_i^{t+1}$. Agents learn individual policies $\pi_i$ by interacting with the environment using individual, decentralised fixed-size Q-tables.

We first introduce the Q-learning methodology.  Then, the mapping between the RL agent action and the decision variables in \Cref{variables} is presented. Finally, we propose variations on the learning method, with different experience sources, multi-agent structures and reward definitions.

\subsection{Q-Learning}
While any reinforcement learning methodology could be used with the framework proposed in this paper, here we focus on Q-learning, a model-free, off-policy RL methodology.  Its simplicity and proof of convergence make it suited to developing novel learning methodologies in newly defined environments \citep{Vazquez-Canteli2019}. State-actions values $Q(s,a)$ represent the expected value of all future rewards $r_t$ $\forall \ t \in \mathcal{T}$ when taking action $a$ in state $s$ according to policy $\pi$:
\begin{equation}
Q(s,a) \triangleq E^{\pi}{[r_{t} + \gamma r_{t+1} + \gamma^2r_{t+2}...|s_t = s, a_t = a ]}
\end{equation}

\noindent where $\gamma$ is the discount factor setting the relative importance of future rewards. Estimates are refined incrementally as
\begin{equation}
\hat{Q}(s,a)\leftarrow \hat{Q}(s,a) + \alpha\delta
\end{equation}
where $\delta$ is the temporal-difference error,
\begin{equation}
\delta = \left(r_t +\gamma \hat{V}(s^\textrm{next})-\hat{Q}(s,a)\right)
\end{equation}
$\hat{V}$ is the state-value function estimate,
 \begin{equation}
\hat{V}(s) = \max_{a^* \in \mathcal{A}(s)}{\hat{Q}(s,a^*)}
\end{equation}
and $\alpha$ is the learning rate. In this work we use hysteretic learners, i.e. chiefly optimistic learners that use an increase rate superior to the decrease rate in order to reduce oscillations in the learned policy due to actions chosen by other agents \citep{Matignon2012, Matignon2007}. For $\beta < 1$:
\begin{equation}
  \alpha  =
    \begin{cases}
      \alpha_0 & \text{if $\delta > 0$}\\
      \alpha_0\beta & \text{otherwise}\\
    \end{cases}       
\end{equation}

Agents follow an $\epsilon$-greedy policy to balance exploration of different state-action pairs and knowledge exploitation. The greedy action with highest estimated rewards is selected with probability $1-\epsilon$ and random actions otherwise. 
\begin{equation}
\label{eqn:greedy}
  a^*  =
    \begin{cases}
      \argmax_{\ a^* \in \mathcal{A}}\hat{Q}(s,a^*) & \text{if $x \sim U(0,1) > \epsilon$}\\
      a \sim p(a) = \frac{1}{|\mathcal{A}|} \ \forall \ a \ \in \mathcal{A}& \text{otherwise}\\
    \end{cases}       
\end{equation}

Henceforth, we refer to the estimates $\hat{Q}$ and $\hat{V}$ as $Q$ and $V$ to reduce the amount of notation.

\subsection{Agent state}
The agent state is defined by the time-dependent grid cost coefficient $\textrm{C}_\textrm{g}^t$, i.e. the sum of the grid electricity price and the product of the carbon intensity of the generation mix at time $t$ and the social cost of carbon.

To convert the RL policy action into local decisions, the agent also requires information on their current PV generation, battery level, flexible loads and indoor air temperature, as described below in \Cref{sec:agentdecision}.

\subsection{Agent action}\label{sec:agentdecision}
% this figures is from the DPhilLog slides section 27.04.20
\begin{figure*}[!t]
\begin{center}
\includegraphics[width=0.7\textwidth]{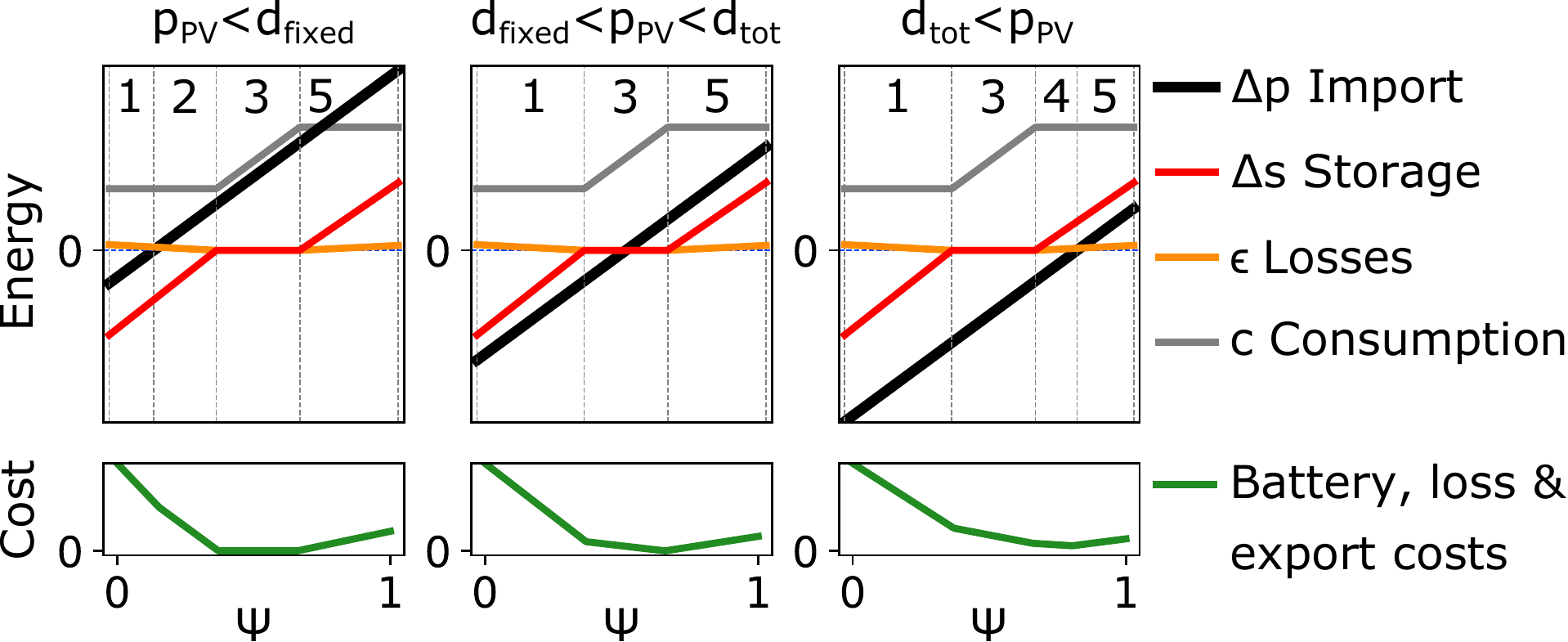}

\end{center}
\caption{Decision variable $\psi$. Sections 1-5 denote the trade-off regimes described in \Cref{sec:agentdecision}. At each step, the fixed requirements for loads, heat and upcoming EV trips are first met. The $\psi$ decision then applies to the remaining flexibility, from maximal energy exports (full use of flexibility) at $\psi = 0$, to maximal energy imports (no use of flexibility) at $\psi = 1$. $\textrm{d}_\textrm{tot}$ and $\textrm{d}_\textrm{fixed}$ are the sum of household and heating loads with and without their flexible component. If fixed loads cannot be fully met by PV energy,  the residual is met by storage and imports (2). If there is additional PV energy after meeting all loads, it can be stored or exported (4).}
\label{fig:mu}
\end{figure*}

Large action spaces compound the curse of dimensionality in Q-learning and waste exploration resources \citep{Powell2011}. At each time step, the decision variables in \Cref{system} controlling the flows in and out of the battery$b_{\textrm{in},i}^t$ and $b_{\textrm{out},i}^t$, the electric heating consumption $h_i^t$ and the prosumer consumption $c_i^t$ for household $i$ are therefore synthesised into a single variable $\psi \in [0,1]$ controlling the use of available local flexibility. \Cref{fig:mu} shows how consumption (for domestic loads and heat), imports and storage change with $\psi$.

At each step, the fixed requirements for loads, heat and upcoming EV trips are first met. The $\psi$ decision then applies to the remaining flexibility. In conditions deemed optimal for energy exports $\psi = 0$, all initial storage and residual PV generation is exported and flexible loads are delayed. On the other end, a \emph{passive} agent does not utilise its flexibility and uses the \emph{default} action $\psi = 1$, maximising imports with EVs charged when plugged in and no flexible loads delayed.  Intermediate imports trade-offs are mapped on \Cref{fig:mu}:
 
\begin{enumerate}
\item From exporting all to none of the initial storage $E_i^t$
\item From meeting fixed loads $\textrm{d}_{i,\textrm{fixed}}^t$ with the energy stored to importing the required amount 
\item From no to maximum flexible consumption $\textrm{d}_{i,\textrm{tot}}^t$ 
\item From exporting to storing PV energy $\textrm{p}_{\textrm{PV},i}^t$ remaining after meeting loads
\item From importing no additional energy to filling up the battery to capacity $\overline{\textrm{E}}_i$
\end{enumerate}

Costlier actions incurring battery depreciation, losses and export costs are towards either $\psi$ extreme, only used in highly beneficial situations (convex local costs function in the lower plot of \Cref{fig:mu}). Ranking actions consistently ensures agents do not waste resources trialling sub-optimal combinations of decisions. For example, it is more cost-efficient to first absorb energy imports by consuming flexible loads, and only use the battery (incurring costs) if imports are large.  

Note that although this action space is continuous, it can be discretised into intervals for implementation in Q-learning. 

\subsection{Variations of the learning method}\label{methodologies}
Different experience sources, reward definitions and MARL structures are proposed within the MARL approach. The performance of these combinations of algorithmic possibilities will be assessed in \Cref{results} to inform effective model design.\\

\subsubsection{Experience sources}
In data-driven strategies, the learning is determined by the collected experience.
\begin{itemize}
\item \textbf{Environment exploration}. Traditionally, agents collect experience by interacting with an environment \citep{Sutton1998}.

\item \textbf{Optimisations}. A novel approach collects experience from optimisations. Learning from entities with more knowledge or using knowledge more effectively than randomly exploring agents has previously been proposed, as with agents \say{mimicking} humans playing video games \citep{Grandmaster}. Similarly, agents learn from convex \say{omniscient} optimisations on historical data with perfect knowledge of current and future variables.  This experience is then used under partial observability and control for stable coordination between prosumers at scale. Note in this case that, although the MARL learning and implementation are model-free, a model of the system is used to run the convex optimisation and produce experience to learn from.  A standard convex optimiser uses the same data that would be used to populate the environment explorations but solves over the whole day-horizon with perfect knowledge of all variables using the problem description in \Cref{system}. Then, at each time step, the system variables are translated into equivalent RL $\{s_t,a_t,r_t, s_{t+1}\}$ tuples for each agent, which are used to update the policies in the same way as for standard Q-learning as presented below.\\
\end{itemize}
 
\subsubsection{MARL structures} Both the centralised and decentralised structures proposed use fixed-size $|\mathcal{S}| \times |\mathcal{A}|$ Q-tables corresponding to individual state-action pairs. The size of a global Q-table referencing all possible combinations of states and actions would grow exponentially with the number of agents. This would limit scalability due to memory limitations and exploration time requirements. Moreover, as strategies proposed in this paper are privacy-preserving, only local state-action pairs are used for individual action selection, wasting the level of detail of a global Q-table.

\begin{itemize}
\item \textbf{Distributed learning}. Each agent $i$ learns its $Q_i$ table with its own experience. No information is shared between agents. 

\item \textbf{Centralised learning}. A single table $Q_\textrm{c}$ uses experience from all agents during pre-learning. All agents use the centrally learned policy for decentralised implementation.
\end{itemize}

\subsubsection{Reward definitions}
The reward definition is central to learning as its maximisation forms the basis for incrementally altering the policy \citep{Sutton1998}.  Assessing the impact of individual actions on global rewards accurately is key to the effective coordination of a large number of prosumers. In the following,  the Q-tables $Q^0$, $Q^\textrm{diff}$,$Q^\textrm{A}$ and $Q^\textrm{count}$ may be either agent-specific $Q_i$ or centralised $Q_\textrm{c}$ based on the MARL structure. We proposed four variations of the Q-table update rule for each experience step tuple collected $(s_i^t, a_i^t, r^t,s_i^{t+1})$.
 \begin{equation}
Q(s_i^t, a_i^t)\leftarrow Q(s_i^t,a_i^t) + \alpha \delta
\end{equation}
\begin{itemize}
\item \textbf{Total reward}. The instantaneous total system reward $r^t = \hat{F}_t$ is used to update the Q-table $Q^0$.
\begin{equation}
\delta = r^t + \gamma V^0(s_i^{t+1}) - Q^0(s_i^t, a_i^t)
\end{equation}
\item \textbf{Marginal reward}. The difference in total instant rewards $r^t$ between that if agent $i$ selects the greedy action and that if it selects the default action is used to update $Q^\textrm{diff}$ \citep{Wolpert2002}. The default action $a_\textrm{default}$ corresponds to $\psi = 1$, where no flexibility is used. The default reward $r^t_{a_{i}=a_\textrm{default}}$, where all agents perform their greedy action apart from agent $i$ which performs the default action, is obtained by an additional simulation.
\begin{equation}
\delta = \left(r^t - r^t_{a_{i}=a_\textrm{default}}\right) + \gamma V^\textrm{diff}(s_i^{t+1}) - Q^\textrm{diff}(s_i^t, a_i^t)
\end{equation}
\item \textbf{Advantage reward}. The post difference between $Q^0$ values when $i$ performs the greedy and the default action is used. This corresponds to the estimated increase in rewards not just instantaneously but over all future states, analogously to in \citep{Foerster2018}. No additional simulations are required as the Q-table values are refined over the normal course of explorations.
\begin{equation}
\delta = \left(Q^0(s_i^t, a_i^t) - Q^0(s_i^t, a_{a_i=a_\textrm{default}})\right) - Q^\textrm{A}(s_i^t, a_i^t)
\end{equation}
\item \textbf{Count}. The Q-table stores the number of times each state-action pair is selected by the optimiser. 
 \begin{equation}
\alpha\delta = 1
\end{equation}
\end{itemize}

\section{Input Data}\label{data}

\begin{table*}[!t]
   \newcommand{\smaller}{\rule[-10pt]{0pt}{20pt}}
   \newcommand{\smallstrut}{\rule[-5pt]{0pt}{30pt}}
\begin{tabularx}{\textwidth}{ c|m{5cm}|m{5cm}}
\toprule
& Normalised profile  &  Scaling factor \\ 
\hline
   PV \smaller &  Randomly selected from current month bank $b_{t+1}=(m)$   & 
   \multirow{2}{=}{\setlength\parskip{\baselineskip}%
   Computed as $\lambda_{t+1} = \lambda_{t} + x$, where $x \sim \Gamma\left(\alpha(b_{t},b_{t+1}),\beta(b_{t},b_{t+1})\right)$} \smaller \\
    \cline{1-2}
    Load \smallstrut &\multirow{2}{=}{\setlength\parskip{\baselineskip}%
Cluster selected based on transition probability $p(k_{t+1} | k_t, w_t, w_{t+1})$ \newline Normalised profile  randomly selected from bank $b_{t+1} = (k_{t+1}, w_{t+1})$} \smallstrut &
 \\ 
\cline{1-1}
\cline{3-3} 
   EV \smallstrut & & Random variable from discrete distribution $p(\lambda_{t+1}|\lambda_t, b_t, b_{t+1})$ \smaller \\
 \bottomrule
\end{tabularx}
  \caption{Markov chain mechanism for selecting behaviour clusters, profiles and scaling factors for input data in subsequent days}
\label{tab:loadnextday}
\end{table*}

This section presents the data that is fed into the model presented in \Cref{system}. Interaction with this data will shape the policies learned through RL \citep{Sutton1998} and should reflect resource intermittency and uncertainty to maximise the expectation of rewards in a robust way without over-fitting. EV demand $\textrm{d}_{\textrm{EV},i}^t$ and availability $\upmu_i^t$, PV production $\textrm{p}_{\textrm{PV},i}^t$ and electricity consumption $\textrm{d}_i^{t}$ are drawn from large representative datasets.

\subsection{Data selection and pre-processing}
Load and PV generation profiles are obtained from the Customer Led Network Revolution (CLNR), a UK-based smart grid demonstration project \citep{TC1a,TC5}, and mobility data from the English National Travel Survey (NTS) \citep{DepartmentforTransport2019}. The NTS does not focus on EVs only and offers a less biased view into the general population's travel pattern than small-scale EV trials data, both due to the smaller volume of data available compared to for generic cars and because the self-selected EV early trial participants may not be representative of patterns once EVs become widely adopted. It is implicitly assumed that electrification will not affect transport patterns \citep{Crozier2018}.

NTS data from ﻿82,455 households from 2002 to 2017 results in 1,272,834 full days of travel profiles. Load and PV data from 11,907 customers between 2011 and 2014 yields 620,702 and 22,670 full days of data, respectively.  Profiles are converted to hourly resolution and single missing points replaced with the figure from the same time the day or week before or after which has the lowest sum of squares of differences between the previous and subsequent point. Tested with available data, this yields absolute errors with mean 0.13 and 0.08 kWh and 99th percentile 1.09 and 0.81 kWh for PV and load data.  PV sources have nominal capacities between 1.35 and 2.02 kWp.

The at home-availability of the vehicles is inferred from the recorded journeys' origin and destination.  EV energy consumption profiles are obtained using representative consumption factors from a tank-to-wheel model proposed in \citep{Crozier2018}, dependent on travel speed and type (rural, urban, motorway). 

\subsection{Markov chain}
\begin{figure*}[!b]
\begin{center}
\includegraphics[width=0.7\textwidth]{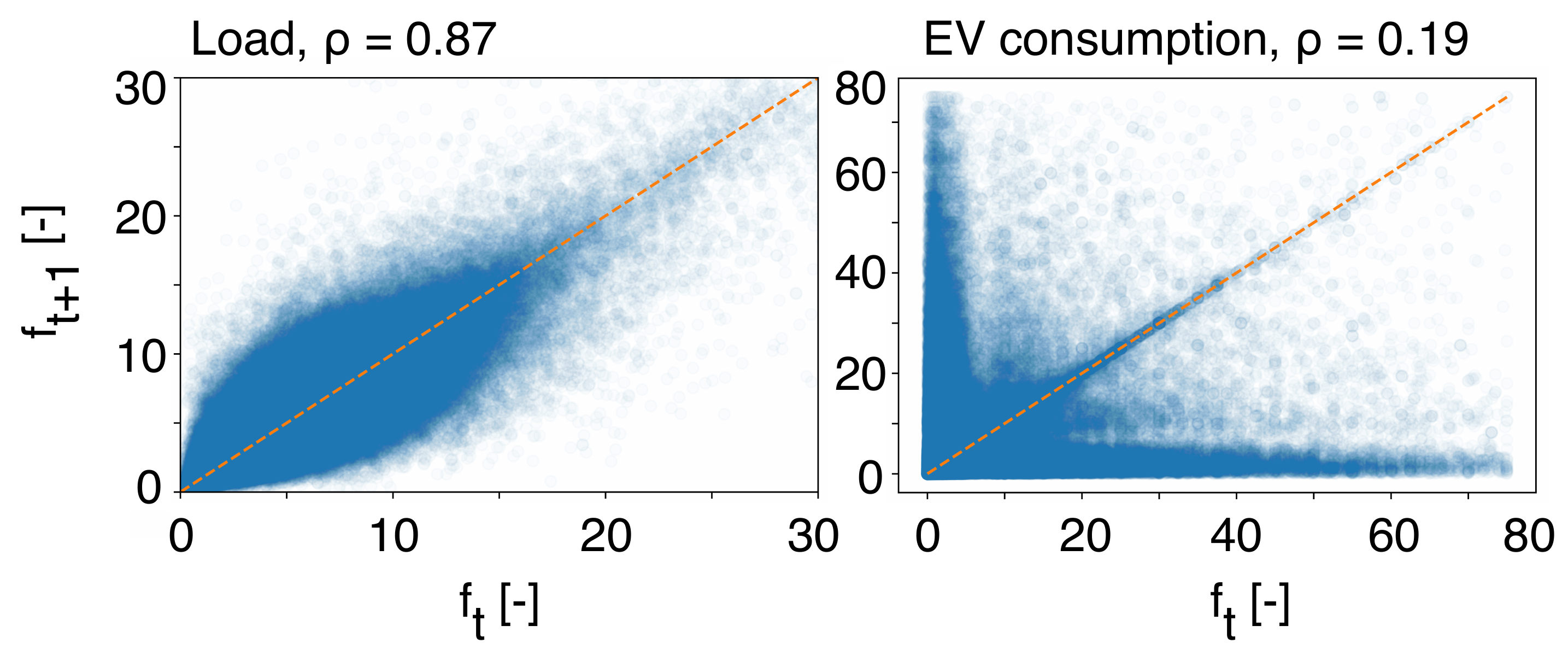}
\end{center}
\caption{Scaling factors for normalised profiles (i.e. total daily loads in kWh) in subsequent days. Linear correlation can be observed for the load profiles,  while more complex patterns are exhibited for EV consumption. $\rho$ is the Pearson correlation coefficient.}
\label{fig:Corr}
\end{figure*}

During learning, agents continuously receive experience to learn from. However,  numerous subsequent days of data are not available for single agents. We design a Markov chain mechanism to feed consistent profiles for successive days, using both consistent scaling factors and behaviour clusters.

Daily profiles for load and travel are normalised such that $\sum_{t=0..24}{x^t}=1$, and clustered using K-means, minimising the within-cluster sum-of-squares \citep{Lloyd1982} in four clusters for both weekday and weekend data (with one for no travel). The features used for load profiles clustering are normalised peak magnitude and time and normalised values over critical time windows, and those for travel are normalised values between 6 am and 10 pm. PV profiles were grouped per month.

Probabilistic Markov chain transition rules are shown in \Cref{tab:loadnextday}. Transition probabilities for clusters $k$ and scaling factors $\lambda$ are obtained from available transitions between subsequent days in the datasets for each week day type $w$ (week day or weekend day). \Cref{fig:Corr} shows that subsequent PV and load scaling factors follow strong linear correlation, with the residuals of the perfect correlation following gamma distributions with zero mean, whereas EV load scaling factors follow more complex patterns, so transitions probabilities are computed between 50 discrete intervals.

\section{Case study results and discussion}\label{results}
This section compares the performance of the residential flexibility coordination strategies presented in \Cref{RLSection} to baseline and upper bound scenarios for increasing numbers of prosumers. The performance of traditionally used MARL strategies drops at scale, while that of the novel optimisation-based methodology using marginal rewards is maintained.

\subsection{Set-up}
The MARL algorithm is trained in off-line simulations using historical data prior to online implementation. This means agents do not trial unsuccessful actions with real-life impacts during learning. Moreover, the computation burden is taken prior to implementation, while prosumers only apply pre-learned policies, avoiding the computational challenges of large-scale real-time control. 

The learning occurs over 50 epochs consisting of an exploration, an update and an evaluation phase. First,  the environment is explored over two training episodes of duration $|\mathcal{T}| = 24$ hours.  Learning in batches of multiple episodes helps stabilise learning in the stochastic environment.  Then,  Q-tables are updated based on the rules presented in \Cref{methodologies}. Finally, an evaluation is performed using a deterministic greedy policy on new evaluation data.  Ten repetitions are performed such that the learning may be assessed over different trajectories.

The Social Cost of Carbon is set at 70 £/tCO$_2$, consistent with the UK 2030 target \citep{Hirst2018}. Weather \citep{WeatherWunderground2020}, electricity time-of-use prices \citep{OctopusEnergy2019} and grid carbon intensity \citep{NationalGridESO2020} are from January 2020, where relevant specified for London, UK. The low solar heat gains in January are neglected \citep{Brown2020}. Other relevant parameters for the case studies are listed in \Cref{app:inputs}.

As performed on a Intel(R) Core(TM) i7-9800X CPU @ 3.80GHz, computation time for a learning trajectory is $2^\prime45^{\prime\prime}$ for one agent and $97^\prime5^{\prime\prime}$ for 30 agents, including evaluation points. The policy can then be directly applied at the household level during operation.

Case study results using different experience sources, reward definitions and MARL structures are presented in \Cref{fig:results}. Acronyms for each strategy are tabulated in the legend. Positive values denote savings relative to a baseline scenario where all agents are passive, i.e. not using their flexibility with EVs charged immediately and no flexible loads delayed. As the Q-learning policies are first initialised with zero values, in the first epoch of learning completely random action values are chosen, which provides rewards far below the baseline. As agents collect experience and update their policies at each epoch, improved policies are learned, some of which are able to outperform the baseline. An upper bound is provided by results from \say{omniscient} convex optimisations, which are however not achievable in practice for three main reasons. Firstly, they use perfect knowledge of all the environment variables in the present and future, despite uncertainty in renewable generation, mix of the grid, and customer behaviour. Optimisation with inaccurate data would lead to suboptimal results. Secondly, prosumers may not be willing to yield their data and direct control to an external entity. Finally, central optimisations become computationally expensive for real-time control of large numbers of prosumers.

\subsection{Results}
\begin{figure*}[t!]
\begin{center}
\includegraphics[width=\textwidth]{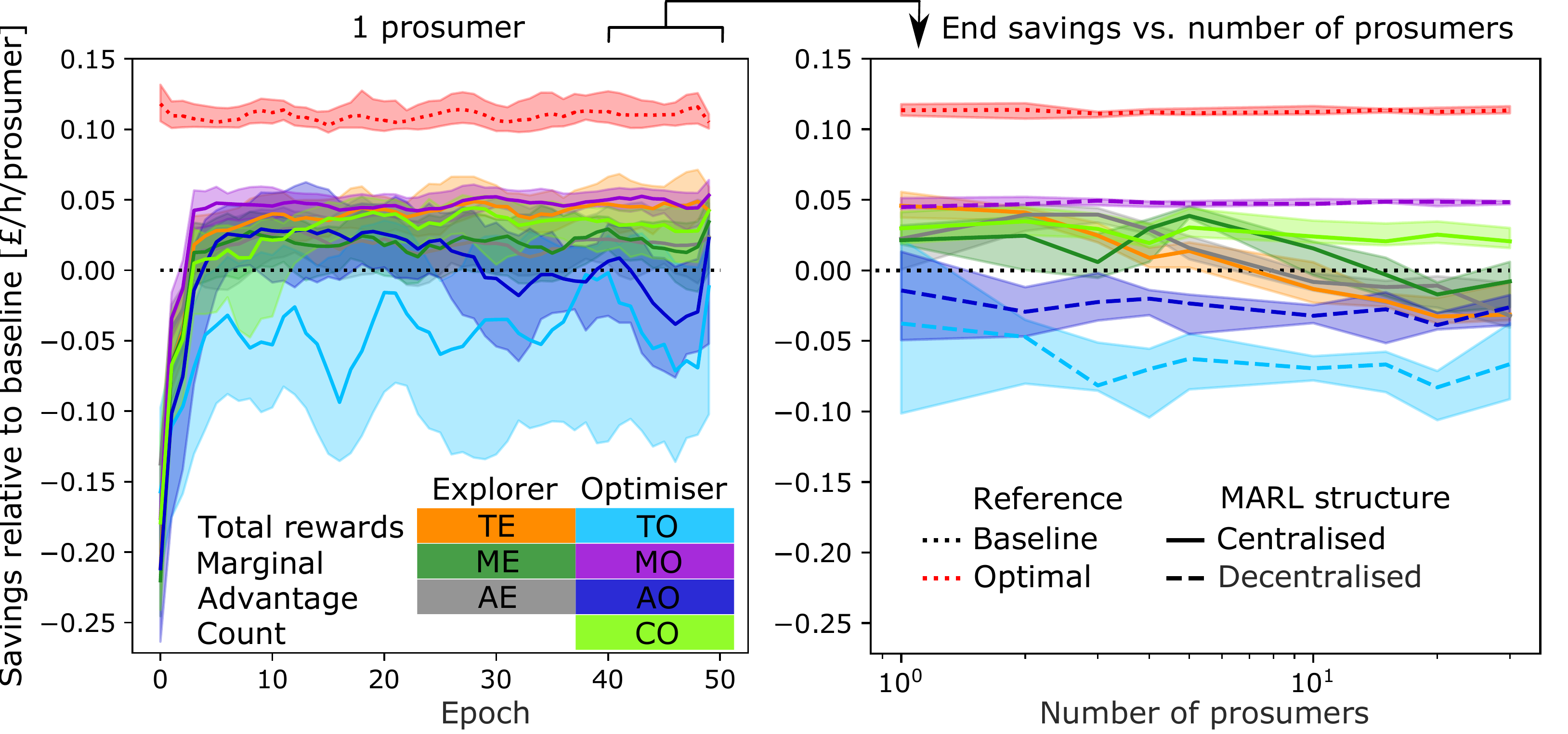}

\end{center}
\caption{The left-hand side plot shows the five-epoch moving average of evaluation rewards relative to baseline rewards for a single prosumer. The right-hand side plot shows the mean of the final 10 evaluations against the number of prosumers.  Lines show median values and shaded areas the 25th and 75th percentiles over the 10 repetitions. The best-performing MARL structure is displayed for each exploration source and reward definition pair.  The performance of the baseline MARL algorithm (TE, orange) drops as the number of concurrently learning agents in the stochastic environment increases; the best-performing alternative algorithm proposed (MO, purple) maintains high performance at scale.}
\label{fig:results}
\end{figure*}

Results presented in \Cref{fig:results} show that only the algorithms learning from optimisations maintained stable coordination performance at scale, while the performance of traditionally used MARL algorithms would drop in this context of stochasticity and partial observation.  The optimisation-based algorithm which uses marginal rewards (MO) performed best.  We further elaborate on the results in the subsections below.

\subsubsection{Environment exploration-based learning}
The centralised MARL structure is favoured for environment exploration-based learning (continuous lines in \Cref{fig:results}).  A single policy uses experience collected by all agents, rather than each agent learning from their own experience only.  

\Cref{fig:results} shows that environment exploration-based MARL using total rewards (TE, orange), the baseline MARL framework, exhibits a high performance for a single agent. However, savings drop as the number of cooperating agents increases, down to around zero from ten agents.  Coordination challenges arise for independent learners to isolate the contribution of their actions to total rewards from the stochasticity of the environment, compounded by other simultaneously learning agents' random explorations, and the non-stationarity of their on-policy behaviour \citep{Matignon2012}.  

Using advantage rewards (AE, grey), based on estimates of the long-term value of actions relative to that of the baseline action, yields superior results beyond two agents. However, as AE uses the total reward $Q^0$-table as an intermediary step, results similarly drops for increasing numbers of agents.

Using marginal rewards (ME, dark green), the value of each agent's action relative to the baseline action is singled out immediately by an additional simulation and used as a reward at each time step.  This improves the performance relative to TE and AE for five agents and more, though still with declining performance as the number of agents increases.

\subsubsection{Optimisation-based learning}
Optimisation-based learning generally favours the distributed MARL structure, with agents able to converge to distinct compatible policies (dashed lines in \Cref{fig:results}).

Comparing trajectories in \Cref{fig:results},  learning from the total rewards obtained by an optimiser (TO, light blue) yields lower savings than when using environment explorations (TE). The learned policies yield negative savings, i.e. would provide worse outcomes than inflexible agents. The omniscient optimiser takes precise, extreme decisions thanks to its perfect knowledge of all current and future system variables, importing at very high $\psi$ values when it is optimal to do so.  RL algorithms on the other hand are used under partial observability, aiming for actions that statistically perform well under uncertainty.  Agents independently picking TO-based decisive actions in a stochastic environment do not yield optimal outcomes. Assessing the long-term advantage of actions from optimisations (AO, dark blue) follows a similar trend, whilst providing marginally superior savings relative to TO.

Optimisation-based learning using marginal rewards (MO, purple) offers the highest savings as the additional baseline simulations are best able to isolate the contribution of individual actions from variations caused by both the environment and other agents.  When increasing the number of agents, the strategy is able to learn from optimal, stable, consistently behaving agents. Savings of 6.18p per agent per hour, or £45.11 per agent per month are obtained on average for 30 agents, corresponding to a 33.7\% reduction from baseline costs.  65.9\% of savings stem from reduced battery depreciation, 20.32\% from distribution grid congestion, 11.1\% from grid energy, and 2.7\% from greenhouse gas emissions.

The count-based strategy learning from optimisations (CO, light green) seeks to reproduce the state-action patterns of the omniscient optimiser with perfect knowledge of system variables and perfect control of agents for local decision-making under partial observability.  It provides results lower than the high performances of MO, though with a stable performance at scale. Savings of £21.09 per agent per month on average for 30 agents are obtained. The battery and distribution grid costs increase by an equivalent of 6.0\% and 7.7\% of total savings respectively, while grid energy and greenhouse gas emissions costs reductions represent 59.7\% and 54.0\% of total savings.

Both the MO and CO strategies exhibit stable performance at scale, though converging to different types of policy. The MO policy saves more by smoothing out the charging and distribution grid utilisation profiles despite smaller savings in imports and emissions costs, while CO derives a larger advantage from the grid price differentials in grid imports, though with higher battery and distribution grid costs. The weight applied to each of those competing objectives in the objective function directly impacts the policies that are learned. Examples of how the individual home energy management system decision variables (heating, energy consumption, battery charging) vary based on the controller are illustrated in \Cref{app:example_case_study}.

Overall, the new class of optimisation-based learning performs significantly better across different numbers of prosumers, with higher savings and lower inter-quartile range than environment-based learning at scale. This superior performance requires computations to run optimisations on historical data, and to perform baseline simulations to compute marginal rewards, though computational time for pre-learning is not strictly a limiting factor as it is performed off-line ahead of implementation. 

A fundamental challenge in MARL has been the trade-off between fully centralised value functions, which are impractical for more than a handful of agents, or, in a more straightforward approach, independent learning of individual action-value functions by each agent in independent Q-learning (IQL) \citep{Tan1993}. However, an ongoing issue with this approach has been that of convergence at scale, as agents do not have explicit representations of interactions between agents, and each agent’s learning is confounded by the learning and exploration of others \citep{Rashid2020}. As shown in \Cref{fig:results}, the Pareto selection, non-stationarity and stochasticity issues presented in \Cref{sec:gapanalysis} have prevented environment exploration-based learners from achieving successful MARL cooperation at scale for agents under partial observability in a stochastic environment. This case study of coordinated residential energy management shows that the novel combination of marginal rewards, which help agents isolate their marginal contribution to total rewards, and the learning from results of convex optimisations, where agents learn successful policy equilibriums from omniscient, stable, and consistent solutions, offer significant improvements on these scalability and convergence issues.  

\section{Conclusion}\label{conclusion}
In this paper, a novel class of strategies has addressed the scalability issue of residential energy flexibility coordination in a cost-efficient and privacy-preserving manner.  The combination of off-line optimisations with multi-agent reinforcement learning provides high, stable coordination performance at scale.

We identified in the literature that the concept of RL-based implicit energy coordination, where energy prosumers cooperate towards global objectives based on local information only, had been under-researched beyond frequency droop control with limited number of agents. The scalability of such methods was identified as a key gap that we have sought to bridge. The novel coordination mechanism proposed in this paper thus satisfies the criteria for successful residential energy coordination set out in the introduction, as tested with large banks of real data in the case studies:

\begin{itemize}

\item Computational scalability: The scalability of traditional learning algorithms is significantly improved thanks to fixed-size Q-tables to avoid the curse of dimensionality, so that policies can be learned for larger number of agents. The proposed method does not require expensive communication and control appliances at the prosumer level, as pre-learned policies are directly applied with no further communication and no exponential time real-time optimisations needed. This is a crucial benefit for applications with physical limitations in hardware availability and processing time.  

\item Performance scalability: The coordination performance remains high for increasing numbers of prosumers despite the challenges of partial observability,  environment stochasticity and concurrently learning of agents, thanks to learning from the results of global omniscient optimisations on historical data, and to rewards signals that isolate individual contributions to global rewards. Significant value of £45.11 per agent per month was obtained in the presented case study for 30 agents,  thanks to savings in energy,  prosumer storage and societal greenhouse gas emissions-related costs. Those savings do not drop with increasing number of agents, as opposed to with standard MARL approaches.

\item Acceptability: The approach does not rely on sharing of personal data, thermal discomfort, or hindrance/delay of activities, and the appliances are controlled locally. This cost-efficient and privacy-preserving implicit coordination approach could help integrate distributed energy resources such as residential energy, otherwise excluded from energy systems' flexibility management. 

\end{itemize}

Important future work is a more detailed assessment of the impacts of the coordination strategies on power flows, as well as an evaluation of the generalisation and adaptability potential of policies when used by other households or if household characteristics change over time.  Moreover, while all agents readily reduce individual costs through participation in the framework, further game-theoretic tools could be used to design a post-operation reward scheme.

\section*{Acknowledgement}
This work was supported by the Saven European Scholarship and by the UK Research and Innovation and the Engineering and Physical Sciences Research Council (award references EP/S000887/1, EP/S031901/1, and EP/T028564/1).

\end{multicols}

\newpage
\begin{appendices}
\section{Nomenclature}\label{app:nomenclature}

\begin{longtable}{p{.18\textwidth} p{.82\textwidth}}
& Acronyms\\
  \hline
AE  & MARL with advantage rewards and exploration-based learning \\ 
AO  & MARL using advantage rewards and optimisation-based learning \\ 
AI & artificial intelligence \\
CLNR & customer-led network revolution \\ 
CO  & MARL using count rewards and optimisation-based learning \\ 
ME  & MARL using marginal rewards and exploration-based learning \\ 
MO  & MARL using marginal rewards and optimisation-based learning \\ 
Dec-POMDP  & decentralised partially observable Markov decision process \\ 
DER & distributed energy resource \\
DR & demand response \\
EV & electric vehicle \\
MARL & multi-agent reinforcement learning \\
NTS & national travel survey \\
PV & photovoltaic \\ 
RL & reinforcement learning \\
TE  & MARL using total rewards and exploration-based learning \\ 
TO  & MARL using total rewards and optimisation-based learning \\ 
UK & United Kingdom \\
  \hline
   & \\ 
   
  & Variables \\
  \hline
$b_\textrm{in}$ & charge into the battery [kWh] \\
$\overline{\textrm{b}_\textrm{in}}$ & maximum charge into the battery [kWh] \\ 
$b_\textrm{out}$ & discharge out of the battery [kWh] \\ 
$c$ & household consumption [kWh] \\ 
$\hat{c}$ & partial consumption for load type and time demanded [kWh] \\ 
$c_\textrm{d}$ & distribution cost [£] \\ 
$\textrm{C}_\textrm{d}$ & distribution charge [£/kWh] \\ 
$c_\textrm{g}$ & grid cost [£] \\ 
$\textrm{C}_\textrm{g}$ & grid cost coefficient [£/kWh] \\ 
$c_\textrm{s}$ & storage cost [£] \\ 
$\textrm{C}_\textrm{s}$ & battery depreciation coefficient [£/kWh] \\ 
$\textrm{d}$ & household demand [kWh] \\ 
$\textrm{d}_\textrm{EV}$ & electric vehicle demand [kWh] \\
$d_\textrm{fixed}$ & sum of non-flexible household and heating loads [kWh] \\ 
$d_\textrm{tot}$ & sum of household all heating loads [kWh] \\ 
$E$ & battery energy level [kWh] \\ 
$E_0$ & initial battery energy level [kWh] \\ 
$\underline{\textrm{E}}$ & minimum battery energy level [kWh] \\
$\overline{\textrm{E}}$ & maximum battery energy level [kWh] \\ 
$\textrm{f}$ & flexibility boolean \\
$F$ & objective function [£] \\
$\hat{F}$ & share of objective function for given time step \\
$g$ & total grid import to the group of prosumers [kWh] \\
$h$ & heating energy consumption [kWh] \\
$k$ & behaviour cluster for transport or household consumption profile \\
$p$ & prosumer import [kWh] \\ 
$\textrm{p}_\textrm{PV}$ & PV generation [kWh] \\ 
$Q$ & Q value [£] \\
$\hat{Q}$ & Q value estimate [£] \\
$r$ & global reward [£] \\
$\textrm{R}$ & average resistance between the main grid and the prosumers [$\Omega$]\\
$\textrm{T}_\textrm{e}$ & external temperature [$^o$C]\\
$\textrm{T}_\textrm{m}$ & building mass temperature [$^o$C]\\
$\textrm{T}_\textrm{air}$ & indoor air temperature [$^o$C]\\
$\underline{\textrm{T}}$ & minimum indoor air temperature [$^o$C]\\
$\overline{\textrm{T}}$ & maximum indoor air temperature [$^o$C]\\
$U$ & uniform distribution function\\
$\textrm{V}$ & nominal root mean square grid voltage [V]\\
$\hat{V}$ & state-value estimate [£] \\
  \hline
  &\\ 
  
 & Greek letters\\
  \hline
$\alpha_0$ & base learning rate [-] \\ 
$\alpha$ & learning rate [-] \\ 
$\beta$ & hysteretic learning rate reduction factor [-] \\ 
$\gamma$ & discount factor [-] \\ 
$\delta$ & loss [£] \\ 
$\epsilon_\textrm{ch}$ & battery charging losses [kWh] \\ 
$\epsilon_\textrm{dis}$ & battery discharging losses [kWh] \\ 
$\epsilon$ & share of random action selection during exploration [-] \\ 
$\upeta_\textrm{ch}$ & battery charging efficiency [-] \\ 
$\upeta_\textrm{dis}$ & battery discharging efficiency [-] \\ 
$\upkappa$ & matrix of heating model coefficients \\ 
$\lambda$ & scaling factor for transport or household consumption profile [kWh] \\
$\upmu$ & electric vehicle availability boolean \\ 
$\pi$ & policy \\ 
$\upphi$ & solar heat flow rate [J.s$^{-1}$] \\ 
$\psi$ & local flexibility use decision variable [-] \\ 

  \hline
  & \\ 
  & Indexes \\ 
  \hline
 $a$ & action\\
$i$ & prosumer \\
$s$ & state\\
$t$ & time step  \\
$t_\textrm{C}$ & consumption time step  \\
$t_\textrm{D}$ & demand time step  \\
$w$ & day type (week day or weekend day)\\

  \hline
  & \\ 
  
 & Sets \\ 
  \hline
$\mathcal{A}$ & set of actions \\
$\mathcal{T}$ & set of time steps\\
$\mathcal{P}$ & set of prosumers\\
$\mathcal{S}$ & set of states\\

  \bottomrule
  \caption{Nomenclature}
\label{tab:nomenclature}
\end{longtable}

\newpage

\section{Case study input data}\label{app:inputs}
\begin{itemize}
\item Learning parameters: The depreciation, learning and exploration rates are $\gamma=$ 0.99, $\alpha_0 =$ 0.01 and $\epsilon =$ 0.5.  The hysteretic learning rate reduction parameter for negative errors is $\beta=0.5$. The states are defined by three uniform grid cost intervals for each day. The action space is discretised in 10 equal $\psi$ intervals.  
\item Battery: $\upeta_\textrm{ch} = \upeta_\textrm{dis} = \sqrt{\upeta_\textrm{round trip}}$ \citep{HOMEREnergy2020},  where $\upeta_\textrm{round trip} = 0.87$ \citep{Schram2020},  capacity $\overline{\textrm{E}} = 75$ kWh, max. charging rate $\overline{\textrm{b}_\textrm{in}} = 22$ kW, depreciation $\textrm{C}_\textrm{s} = 20$ USD/MWh-throughput \citep{Multiclass}, initial and min. charge $\textrm{E}_0 = 0.5\overline{\textrm{E}}$ and $\underline{\textrm{E}} = 0.1\overline{\textrm{E}}$.
\item Grid: nominal voltage $\textrm{V} = 415$ [V], average resistance to prosumers $\textrm{R} = 0.084$ [$\Omega$] \citep{Multiclass}. 
\item Flexible loads: 10\% deferrable for up to $\textrm{n}_\textrm{flex} = 5$ hours.
\item Heating: housing of 76 m$^2$, 2.4m height. Comfort temperature 20$^o$C between 7-10am and 5-10pm, setback 16$^o$C. Variations of 3$^o$C acceptable. U-values from \citep{Becker2018}, other heating inputs from \citep{ISO2007,BRE2014,BSI2009}. Pre-heating up to five hours in advance. Coefficients after re-arranging:
\begin{equation}
{\upkappa =
 \begin{bmatrix}
 6.84\mathrm{e}{-2}, 9.08\mathrm{e}{-1}, 9.15\mathrm{e}{-2}, 2.62\mathrm{e}{-4}, 2.52\mathrm{e}{-1}\\[0.5pt]
2.40\mathrm{e}{-1}, 8.80\mathrm{e}{-1}, 1.20\mathrm{e}{-1}, 3.46\mathrm{e}{-4}, 1.46
\end{bmatrix}}
\end{equation}
\item EV consumption factors [kWh/10km]: 2.25 for motorway, 1.62 for urban and 1.36 for rural travel \citep{Crozier2018}.
\item Distribution network export charge: 0.01 £/kWh

\end{itemize}

\newpage
%\addtocounter{page}{-17}

\section{Heating model}\label{app:heating}
We use the simple hourly method heating model laid out in \citep{ISO2007}.

The input data used in the heating model in this paper is tabulated below. Note that these heating model and input data are meant as a generic building example that can be used to test the relative performances of the MARL coordination algorithms in the case study in \Cref{results}. Models and parameters used for the detailed study of a specific building should be validated with experimental data.

\begin{table}[H]
\begin{tabularx}{\textwidth}{ m{1.2cm} m{5.5cm}  l l l }
\toprule
Symbol & Definition & Value & Unit & Reference \\ 
$A_\textrm{d}$ & door area & 1.4 $\times$ 2 & m$^2$ & \\
$A_\textrm{f}$ & floor area & 76 & m$^2$ & \\
$A_\textrm{wd}$ & window area & 1.4$\times$1.4 & m$^2$ & \\
$e$ & shielding coefficient & 0.03 & - & \citep{BSI2009} \\ 
$h$ & height of rooms & 2.4 & m & \\
$h_\textrm{is}$ & heat transfer coefficient between the air node $\theta_\textrm{air} $ and the surface node $\theta_\textrm{s} $ & 3.45 & W.m$^{-2}$.K$^{-1}$ & \citep{ISO2007} \\ 
$h_\textrm{ms}$ &  heat transfer coefficient between nodes m and s & 9.1 & W.m$^{-2}$.K$^{-1}$ & \citep{ISO2007} \\ 
$k_\textrm{party}$ & fraction of floor space that is party floor rather than on ground, for one-storey building & 0.5 & [-] & \\
$n_\textrm{min}$ & minimum external air exchange rate per hour for a habitable room & 0.5 & h$^{-1}$ & \citep{BSI2009} \\ 
$n_{50}$ & Air exchange rate resulting from a pressure difference of 50 Pa between the inside and the outside of the building, including the effects of air inlets, medium construction family dwelling & 6 & h$^{-1}$ & \citep{BSI2009} \\ 
$U_\textrm{g}$ & U-value for ground (old build) & 1.0 & W.m$^{-2}$.K$^{-1}$ & \citep{Becker2018} \\
$U_\textrm{r}$ & U-value for roof (old build) & 1.0 & W.m$^{-2}$.K$^{-1}$ & \citep{Becker2018} \\
$U_\textrm{w}$ & U-value for walls, ceiling against outside (old build) & 1.5 & W.m$^{-2}$.K$^{-1}$ & \citep{Becker2018} \\
$U_\textrm{wd}$ & U-value for windows (old build) & 4.3 & W.m$^{-2}$.K$^{-1}$ & \citep{Becker2018} \\
$\epsilon$ & height correction factor & 1 & - & \citep{BSI2009} \\                      
$\Lambda_\textrm{at}$ & dimensionless ratio between the internal surfaces area and the floor area & 4.5 & [-] & \citep{ISO2007} \\ 
$\tau$  & time step & 3600 & s & \\          
 \bottomrule
\end{tabularx}
  \caption{Input parameters to the heating model}
\label{tab:heat_inputs}
\end{table}

We obtain the following intermediate parameter values:
\begin{itemize}
\item Effective mass area $A_\textrm{m}$ [m$^{2}$] for a medium-class building \citep{ISO2007}:
\begin{equation}
A_\textrm{m} = 2.5A_\textrm{f}
\end{equation}

\item Internal heat capacity of the building zone for medium-class building  J.K$^{-1}$ \citep{ISO2007}:
\begin{equation}
C_\textrm{m} = 165,000 A_\textrm{f}
\end{equation}  

\item Area of all surfaces facing the building zone $A_\textrm{tot}$  [m$^{2}$] \citep{ISO2007}:
\begin{equation}
A_\textrm{tot} = \Lambda_\textrm{at}A_\textrm{f}
\end{equation}

\item The coupling conductance [W.K$^{-1}$] \citep{ISO2007}:
\begin{equation}
H_\textrm{tr,is} = h_\textrm{is}A_\textrm{tot}
\end{equation}

\item The coupling conductance between nodes m and s [W.K$^{-1}$] \citep{ISO2007}:
\begin{equation}
H_\textrm{tr,ms} = h_\textrm{ms}A_\textrm{m}
\end{equation}
        
\item Wall area (excluding windows and doors)
\begin{equation}
A_\textrm{w} = 4\sqrt{A_f}  h-8 A_\textrm{wd}-A_\textrm{d}
\end{equation}

\item The thermal transmission coefficient of walls [W.K$^{-1}$] \citep{ISO2007}:
\begin{equation}
H_\textrm{tr,w} =  A_\textrm{w} U_\textrm{w}
\end{equation}

\item The thermal transmission coefficienf of the roof [W.K$^{-1}$] \citep{ISO2007}:
\begin{equation}
H_\textrm{tr,r} =  A_\textrm{f} U_\textrm{r}
\end{equation}

\item The thermal transmission coefficient of the floor [W.K$^{-1}$] \citep{ISO2007}:
\begin{equation}
H_\textrm{tr,f} =  A_\textrm{f} (1-k_\textrm{party})U_\textrm{g}
\end{equation}

\item The heat transfer coefficient for opaque elements $H_\textrm{tr,op}$  [W.K$^{-1}$] \citep{ISO2007}:
\begin{equation}
H_\textrm{tr,op}  = H_\textrm{tr,w}  + H_\textrm{tr,r}  + H_\textrm{tr,f}
\end{equation}

\item The opaque heat transfer coefficient is split between conductance transfer and $H_\textrm{tr,em}$ \citep{ISO2007}:
\begin{equation}
H_\textrm{tr,em}  = \frac{1}{\frac{1}{H_\textrm{tr,op}} - \frac{1}{H_\textrm{tr,ms}}}              
\end{equation}

\item The thermal transmission coefficienf of windows [W.K$^{-1}$] \citep{ISO2007}:
\begin{equation}
H_\textrm{tr,wd} =  (A_\textrm{wd}+A_\textrm{d}) U_\textrm{wd,eff}
\end{equation}

\item The conditioned air volume [m$^3$]
\begin{equation}
V_\textrm{r} = A_\textrm{f}h
\end{equation}

\item The hygiene minimuim air flow rate of a heated space $V_\textrm{min}$  [m$^3$.h$^{-1}$] \citep{BSI2009}:
\begin{equation}
V_\textrm{min} = n_\textrm{min} V_\textrm{r} 
\end{equation}         

\item The infiltration through building envelope $V_\textrm{inf}$  [m$^3$.h$^{-1}$] \citep{BSI2009}:
\begin{equation}
V_\textrm{inf} = 2 V_\textrm{r} n_{50}e\epsilon
\end{equation}  

\item The air flow rate of heated space [m$^3$.h$^{-1}$]  \citep{BSI2009}:
\begin{equation}
V = \max(V_\textrm{min},V_\textrm{inf} )
\end{equation} 

\item The heat transfer by ventilation $H_\textrm{ve}$ [W.K$^{-1}$] \citep{BSI2009}:
\begin{equation}
H_\textrm{ve} = 0,34 V
\end{equation}

\item The effective window U-value, corrected for the assumed use of curtains [W.m$^{-2}$.K$^{-1}$] \citep{ISO2007}:
\begin{equation}
U_\textrm{wd,eff}=\frac{1}{\frac{1}{U_\textrm{wd}} + 0.04}
\end{equation}

\item Three helper transmission coefficients \citep{ISO2007}:
\begin{equation}
H_\textrm{tr,1} =  \frac{1}{\frac{1}{H_\textrm{ve}}+\frac{1}{H_\textrm{tr,is}}}
\end{equation}\begin{equation}
H_\textrm{tr,2} =  H_\textrm{tr,1}  + H_\textrm{tr,w}
\end{equation}\begin{equation}
H_\textrm{tr,3} =   \frac{1}{\frac{1}{H_\textrm{tr,2}}+\frac{1}{H_\textrm{tr,ms}}}
\end{equation}

\item The heat flow rate from internal heat sources $\Phi_\textrm{int}$ [W] is taken as the sum of the average heat flow rate from appliances $\Phi_\textrm{int,A}$ and occupants $\Phi_\textrm{int,OC}$ \citep{ISO2007}: 
\begin{equation}
 \Phi_\textrm{int} = \Phi_\textrm{int,A} + \Phi_\textrm{int,OC} = 2A_\textrm{f} + 1.5 A_\textrm{f} = 3.5A_\textrm{f}
\end{equation}

\item the part of the heat flow rate from internal heat sources going to the air node $\theta_\textrm{int}$ $\Phi_\textrm{ia}$  [W] \citep{ISO2007}
\begin{equation}
 \Phi_\textrm{ia} = \frac{1}{2} \Phi_\textrm{int}
\end{equation}
\end{itemize}

Given these input parameters, the Crank-Nicholson scheme is defined in \citep{ISO2007} is applied. We seek to find the temperature of the internal air node $\theta_\textrm{air} $ [$^o$C] and of the building mass $\theta_\textrm{m,t} $ at each time step given the heating or cooling power $\Phi_\textrm{HC} $ (positive for heating and negative for cooling), the external air temperature $\theta_\textrm{e}$ [$^o$C] and the heat flow rates from solar heat sources $\Phi_\textrm{sol} $.

The air node temperature $\theta_\textrm{air} $ is given as
\begin{equation}\label{eq:theta_air}
\theta_\textrm{air}   =\frac{H_\textrm{tr,is}  \theta_\textrm{s} + H_\textrm{ve}\theta_\textrm{sup}  + \Phi_\textrm{ia} + \Phi_\textrm{HC} }{H_\textrm{tr,is}   + H_\textrm{ve} }
\end{equation}
Where the surface node temperature $\theta_\textrm{s} $ is defined as:
\begin{equation}\label{eq:theta_s}
\theta_\textrm{s} =\frac{H_\textrm{tr,ms} \theta_\textrm{m} + \Phi_\textrm{st}  +H_\textrm{tr,w} \theta_\textrm{e} +H_\textrm{tr,1} \left(\theta_\textrm{sup}  +\frac{\Phi_\textrm{ia} + \Phi_\textrm{HC} }{H_\textrm{ve} }\right)}{H_\textrm{tr,ms}  + H_\textrm{tr,w}  + H_\textrm{tr,1} }
\end{equation}
The average temperature over the hour of the building mass $\theta_\textrm{m} $:
\begin{equation}\label{eq:theta_m}
\theta_\textrm{m} =\frac{1}{2} (\theta_\textrm{m,t-1} +\theta_\textrm{m,t} )
\end{equation}

\begin{equation}\label{eq:theta_next}
\theta_\textrm{m,t}  = \frac{\theta_\textrm{m,t-1}  \left(\frac{C_\textrm{m}}{\tau} + \frac{1}{2}(H_\textrm{3} + H_\textrm{tr,em})\right) + \Phi_\textrm{mtot} }{\frac{C_\textrm{m}}{\tau} + \frac{1}{2}(H_\textrm{tr,3} +H_\textrm{tr,em})}
\end{equation}

\begin{equation}\label{eq:Phi_mtot}
\Phi_\textrm{mtot}  = \Phi_\textrm{m}  + H_\textrm{tr,em}\theta_\textrm{e}  + H_\textrm{tr,3}  \frac{\Phi_\textrm{st} + H_\textrm{tr,w}\theta_e + H_\textrm{tr,1} \left( \frac{\Phi_\textrm{ia} +\Phi_\textrm{HC} }{H_\textrm{ve}} + \theta_\textrm{sup} \right)}{H_\textrm{tr,2} }
\end{equation}

The part of heat flow rates from internal and solar heat sources going to the internal nodes $\theta_\textrm{s} $ 
\begin{equation}\label{eq:psi_st}
\Phi_\textrm{st}  = \left( 1 - \frac{A_\textrm{m}}{A_\textrm{t}} - \frac{H_\textrm{tr,w}}{9.1 A_\textrm{t}} \right) \left(\frac{1}{2} \Phi_\textrm{int}+\Phi_\textrm{sol}   \right)	
\end{equation}

The part of heat flow rates from internal and solar heat sources going to the internal nodes $\theta_\textrm{m} $ 
\begin{equation}\label{eq:Phi_m}
\Phi_\textrm{m}  = \frac{A_\textrm{m}}{A_\textrm{t}}
\left(\frac{1}{2}\Phi_\textrm{int}+\Phi_\textrm{sol}  \right)
\end{equation}

\begin{equation}\label{eq:theta_sup}
\theta_\textrm{sup}=\theta_\textrm{e}
\end{equation}

We rearrange the equations of this model in order to obtain a linear recursive formulation.
We first define some helper variables:
\begin{equation}\label{eq:A}
A=\frac{C_\textrm{m}}{\tau} + \frac{1}{2}(H_\textrm{tr,3} +H_\textrm{tr,em})
\end{equation}

\begin{equation}\label{eq:B}
B=1 - \frac{A_\textrm{m}}{A_\textrm{t}} - \frac{H_\textrm{tr,w}}{9.1 A_\textrm{t}}
\end{equation}

\begin{equation}\label{eq:C}
C = \frac{B \Phi_\textrm{int}}{2}
\end{equation}

\begin{equation}\label{eq:D}
D=\frac{A_\textrm{m} \Phi_\textrm{int}}{2A_\textrm{t}} + \frac{H_\textrm{tr,3} }{H_\textrm{tr,2} }\left(C+  \frac{H_\textrm{tr,1}\Phi_\textrm{ia} }{H_\textrm{ve}} \right) 
\end{equation}

\begin{equation}\label{eq:E}
E = H_\textrm{tr,em} + \frac{H_\textrm{tr,3} }{H_\textrm{tr,2} }(H_\textrm{tr,w} + H_\textrm{tr,1} ) 
\end{equation}

\begin{equation}\label{eq:F}
H_\textrm{tr,ms}  + H_\textrm{tr,w}  + H_\textrm{tr,1}
\end{equation}

\begin{equation}\label{eq:G}
G=\frac{1}{F} \left(\frac{H_\textrm{tr,ms} a_\textrm{T}}{2}+C+\frac{H_\textrm{tr,1} \Phi_\textrm{ia} }{H_\textrm{ve}} \right)
\end{equation}

\begin{equation}\label{eq:H}
H = \frac{H_\textrm{tr,ms}}{2F} (1+b_\textrm{T} )
\end{equation}

\begin{equation}\label{eq:I}
I = \frac{1}{F}\left(\frac{H_\textrm{tr,ms} c_\textrm{T}}{2}+H_\textrm{tr,w}+H_\textrm{tr,1} \right)
\end{equation}

\begin{equation}\label{eq:J}
J = \frac{1}{F} \left(\frac{H_\textrm{tr,ms} d_\textrm{T}}{2}+B\right)
\end{equation}

\begin{equation}\label{eq:K}
K = \frac{1}{F} \left(\frac{H_\textrm{tr,ms} e_\textrm{T}}{2}+\frac{H_\textrm{tr,1}}{H_\textrm{ve}} \right)
\end{equation}

\begin{equation}\label{eq:aT}
a_\textrm{T} = \frac{D}{A}
\end{equation}

\begin{equation}\label{eq:bT}
b_\textrm{T} = \frac{ \left(\frac{C_\textrm{m}}{\tau} + 0.5(H_\textrm{3} + H_\textrm{tr,em})\right) }{A} 
\end{equation}

\begin{equation}\label{eq:cT}
c_\textrm{T} = \frac{E}{A}
\end{equation}

\begin{equation}\label{eq:dT}
d_\textrm{T} = \frac{\frac{A_\textrm{m}}{A_\textrm{t}} +  \frac{H_\textrm{tr,3} B}{H_\textrm{tr,2} } }{A}
\end{equation}

\begin{equation}\label{eq:eT}
e_\textrm{T}  = \frac{H_\textrm{tr,3} H_\textrm{tr,1}}{H_\textrm{tr,2} H_\textrm{ve}A}
\end{equation}

\begin{equation}\label{eq:a_a}
a_\textrm{air} =  \frac{H_\textrm{tr,is} G + \Phi_\textrm{ia} }{H_\textrm{tr,is}   + H_\textrm{ve}} 
\end{equation}

\begin{equation}\label{eq:b_a}
b_\textrm{air} =  \frac{H_\textrm{tr,is} H}{H_\textrm{tr,is}   + H_\textrm{ve}}
\end{equation}

\begin{equation}\label{eq:c_a}
c_\textrm{air} =  \frac{H_\textrm{tr,is} I + H_\textrm{ve}}{H_\textrm{tr,is}   + H_\textrm{ve}}
\end{equation}

\begin{equation}\label{eq:d_a}
d_\textrm{air} =  \frac{H_\textrm{tr,is} J}{H_\textrm{tr,is}   + H_\textrm{ve}}
\end{equation}

\begin{equation}\label{eq:e_a}
e_\textrm{air} =\frac{H_\textrm{tr,is} K + 1}{H_\textrm{tr,is}   + H_\textrm{ve}}
\end{equation}

\begin{equation}\label{eq:kappa}
\upkappa = \begin{bmatrix}
a_\textrm{T} & b_\textrm{T} & c_\textrm{T} & d_\textrm{T} & e_\textrm{T} \\
a_\textrm{air} & b_\textrm{air} & c_\textrm{air} & d_\textrm{air} & e_\textrm{air} \\
\end{bmatrix}
\end{equation}

Rearranging \cref{eq:B,eq:C,eq:psi_st}:
\begin{equation}\label{eq:psi_st_2}
\Phi_\textrm{st} =C+B\Phi_\textrm{sol}  
\end{equation}

Rearranging \cref{eq:Phi_mtot,eq:Phi_m,eq:psi_st_2,eq:theta_sup}:
\begin{equation}
\begin{split}
\Phi_\textrm{mtot}  = \left( \frac{A_\textrm{m}}{2A_\textrm{t}} \Phi_\textrm{int}+\frac{A_\textrm{m}}{A_\textrm{t}} \Phi_\textrm{sol}    \right) + H_\textrm{tr,em}\theta_\textrm{e}  + \frac{H_\textrm{tr,3} }{H_\textrm{tr,2} }  \left(C+B\Phi_\textrm{sol}  \right)+\\ \frac{H_\textrm{tr,3} }{H_\textrm{tr,2} } H_\textrm{tr,w}\theta_e + \frac{H_\textrm{tr,3} }{H_\textrm{tr,2} }H_\textrm{tr,1} \left( \frac{\Phi_\textrm{ia} +\Phi_\textrm{HC} }{H_\textrm{ve}} + (\theta_\textrm{e} )\right) 
\end{split}
\end{equation}

\begin{equation}
\begin{split}
\Phi_\textrm{mtot}  = \frac{A_\textrm{m} \Phi_\textrm{int}}{2A_\textrm{t}}+\frac{A_\textrm{m}}{A_\textrm{t}} \Phi_\textrm{sol}    + H_\textrm{tr,em}\theta_\textrm{e}  + \frac{H_\textrm{tr,3} C}{H_\textrm{tr,2} } + \frac{H_\textrm{tr,3} B}{H_\textrm{tr,2} } \Phi_\textrm{sol}  + \frac{H_\textrm{tr,3} H_\textrm{tr,w}}{H_\textrm{tr,2} } \theta_e + \\
\frac{H_\textrm{tr,3} H_\textrm{tr,1}\Phi_\textrm{ia} }{H_\textrm{tr,2} H_\textrm{ve}}  + \frac{H_\textrm{tr,3} H_\textrm{tr,1}}{H_\textrm{tr,2} H_\textrm{ve}} \Phi_\textrm{HC}  +\frac{H_\textrm{tr,3} H_\textrm{tr,1}}{H_\textrm{tr,2} }\theta_\textrm{e} 
\end{split}
\end{equation}

\begin{equation}\label{eq:Phi_mtot_2}
\begin{split}
\Phi_\textrm{mtot}  = \left(\frac{A_\textrm{m} \Phi_\textrm{int}}{2A_\textrm{t}} + \frac{H_\textrm{tr,3} }{H_\textrm{tr,2} }\left(C+  \frac{H_\textrm{tr,1}\Phi_\textrm{ia} }{H_\textrm{ve}} \right) \right)
+\left( \frac{A_\textrm{m}}{A_\textrm{t}} +  \frac{H_\textrm{tr,3} B}{H_\textrm{tr,2} } \right) \Phi_\textrm{sol}  \\  
+ \left(H_\textrm{tr,em} + \frac{H_\textrm{tr,3} }{H_\textrm{tr,2} }(H_\textrm{tr,w} + H_\textrm{tr,1} ) \right) \theta_\textrm{e}   + \frac{H_\textrm{tr,3} H_\textrm{tr,1}}{H_\textrm{tr,2} H_\textrm{ve}} \Phi_\textrm{HC}   
\end{split}
\end{equation}

Rearranging \cref{eq:Phi_mtot_2,eq:D,eq:E}:

\begin{equation}\label{eq:Phi_mtot_3}
\Phi_\textrm{mtot}  =D+\left( \frac{A_\textrm{m}}{A_\textrm{t}} +  \frac{H_\textrm{tr,3} B}{H_\textrm{tr,2} } \right) \Phi_\textrm{sol}  
+ E \theta_\textrm{e}   + \frac{H_\textrm{tr,3} H_\textrm{tr,1}}{H_\textrm{tr,2} H_\textrm{ve}} \Phi_\textrm{HC}   
\end{equation}

From \cref{eq:theta_next,eq:A}
\begin{equation}\label{eq:theta_next_2}
\theta_\textrm{m,t}  = \frac{ \left(\frac{C_\textrm{m}}{\tau} + 0.5(H_\textrm{3} + H_\textrm{tr,em})\right) }{A} \theta_\textrm{m,t-1}  + \frac{ \Phi_\textrm{mtot} }{A}
\end{equation}

From \cref{eq:Phi_mtot_3,eq:theta_next_2}
\begin{equation}\label{eq:theta_next_3}
\begin{split}
\theta_\textrm{m,t}  = \frac{ D}{A} +  \frac{ \frac{C_\textrm{m}}{\tau} + 0.5(H_\textrm{3} + H_\textrm{tr,em}) }{A} \theta_\textrm{m,t-1} + \frac{E}{A} \theta_\textrm{e} + \frac{\frac{A_\textrm{m}}{A_\textrm{t}} +  \frac{H_\textrm{tr,3} B}{H_\textrm{tr,2} } }{A} \Phi_\textrm{sol}    +\\  \frac{H_\textrm{tr,3} H_\textrm{tr,1}}{H_\textrm{tr,2} H_\textrm{ve}A} \Phi_\textrm{HC}   
\end{split}
\end{equation}

From \cref{eq:aT,eq:bT,eq:cT,eq:dT,eq:eT,eq:theta_next_3}:
\begin{equation}\label{eq:theta_next_final}
\theta_\textrm{m,t}  = a_\textrm{T} + b_\textrm{T}\theta_\textrm{m,t-1}  + c_\textrm{T}\theta_\textrm{e}  +d_\textrm{T}\Phi_\textrm{sol}   +e_\textrm{T} \Phi_\textrm{HC}   
\end{equation}

From \cref{eq:theta_m,eq:theta_next_final}:
\begin{equation}
\theta_\textrm{m} =\frac{1}{2}\left(\theta_\textrm{m,t-1} +a_\textrm{T}+b_\textrm{T} \theta_\textrm{m,t-1} +c_\textrm{T} \theta_\textrm{e} +d_\textrm{T} \Phi_\textrm{sol}   +e_\textrm{T} \Phi_\textrm{HC} \right)
\end{equation}

\begin{equation}\label{eq:theta_m_2}
\theta_\textrm{m} =\frac{a_\textrm{T}}{2}+\frac{1+b_\textrm{T}}{2} \theta_\textrm{m,t-1} +\frac{c_\textrm{T}}{2} \theta_\textrm{e} +\frac{d_\textrm{T}}{2} \Phi_\textrm{sol}  +\frac{e_\textrm{T}}{2} \Phi_\textrm{HC} 
\end{equation}

From \cref{eq:theta_s,eq:theta_m_2,eq:F,eq:psi_st_2,eq:theta_sup}
\begin{equation}
\begin{split}
\theta_\textrm{s} =\frac{H_\textrm{tr,ms} a_\textrm{T}}{2F}+\frac{H_\textrm{tr,ms}}{F}  \frac{1+b_\textrm{T}}{2} \theta_\textrm{m,t-1} +\frac{H_\textrm{tr,ms} c_\textrm{T}}{2F} \theta_\textrm{e} +\frac{H_\textrm{tr,ms} d_\textrm{T}}{2F} \Phi_\textrm{sol}   + \\ \frac{H_\textrm{tr,ms} e_\textrm{T}}{2F} \Phi_\textrm{HC} +  \frac{C}{F}+\frac{B}{F} \Phi_\textrm{sol}  +\frac{H_\textrm{tr,w}}{F} \theta_\textrm{e} +\frac{H_\textrm{tr,1}}{F} \theta_\textrm{e} +\frac{H_\textrm{tr,1} \Phi_\textrm{ia} }{FH_\textrm{ve} }+\frac{H_\textrm{tr,1}}{FH_\textrm{ve}} \Phi_\textrm{HC} 
\end{split}
\end{equation}

\begin{equation}\label{eq:theta_s_2}
\begin{split}
\theta_\textrm{s} =\frac{1}{F} \left(\frac{H_\textrm{tr,ms} a_\textrm{T}}{2}+C+\frac{H_\textrm{tr,1} \Phi_\textrm{ia} }{H_\textrm{ve}} \right)+\frac{H_\textrm{tr,ms}}{2F} (1+b_\textrm{T} ) \theta_\textrm{m,t-1} +\\  \frac{1}{F}\left(\frac{H_\textrm{tr,ms} c_\textrm{T}}{2}+H_\textrm{tr,w}+H_\textrm{tr,1} \right) \theta_\textrm{e} + \\ \frac{1}{F} \left(\frac{H_\textrm{tr,ms} d_\textrm{T}}{2}+B\right) \Phi_\textrm{sol}  +\frac{1}{F} \left(\frac{H_\textrm{tr,ms} e_\textrm{T}}{2}+\frac{H_\textrm{tr,1}}{H_\textrm{ve}} \right) \Phi_\textrm{HC} 
\end{split}
\end{equation}

From \cref{eq:G,eq:H,eq:I,eq:J,eq:K,eq:theta_s_2}:
\begin{equation}\label{eq:theta_s_3}
\theta_\textrm{s} =G+H \theta_\textrm{m,t-1} +I \theta_\textrm{e} + J \Phi_\textrm{sol}  +K \Phi_\textrm{HC} 
\end{equation}

From \cref{eq:theta_air,eq:theta_s_3,eq:theta_sup}:
\begin{equation}
\begin{split}
\theta_\textrm{air}  = \frac{H_\textrm{tr,is} G}{H_\textrm{tr,is}   + H_\textrm{ve}} +  \frac{H_\textrm{tr,is} H}{H_\textrm{tr,is}   + H_\textrm{ve}}\theta_\textrm{m,t-1}  +  \frac{H_\textrm{tr,is} I}{H_\textrm{tr,is}   + H_\textrm{ve}}\theta_\textrm{e}   + \frac{H_\textrm{tr,is} J}{H_\textrm{tr,is}   + H_\textrm{ve}}\Phi_\textrm{sol}  + \\ \frac{H_\textrm{tr,is} K}{H_\textrm{tr,is}   + H_\textrm{ve}}\Phi_\textrm{HC}  + \frac{H_\textrm{ve}}{H_\textrm{tr,is}   + H_\textrm{ve} }\theta_\textrm{e}  + \frac{\Phi_\textrm{ia} }{H_\textrm{tr,is}   + H_\textrm{ve} }+ \frac{1}{H_\textrm{tr,is}   + H_\textrm{ve} }\Phi_\textrm{HC} 
\end{split}
\end{equation}

\begin{equation}\label{eq:theta_air_2}
\begin{split}
\theta_\textrm{air}  =  \frac{H_\textrm{tr,is} G + \Phi_\textrm{ia} }{H_\textrm{tr,is}   + H_\textrm{ve}} +  \frac{H_\textrm{tr,is} H}{H_\textrm{tr,is}   + H_\textrm{ve}}\theta_\textrm{m,t-1}  +  \frac{H_\textrm{tr,is} I + H_\textrm{ve}}{H_\textrm{tr,is}   + H_\textrm{ve}}\theta_\textrm{e}   + \frac{H_\textrm{tr,is} J}{H_\textrm{tr,is}   + H_\textrm{ve}}\Phi_\textrm{sol}    + \\ \frac{H_\textrm{tr,is} K + 1}{H_\textrm{tr,is}   + H_\textrm{ve}}\Phi_\textrm{HC}  
\end{split}
\end{equation}

From \cref{eq:a_a,eq:b_a,eq:c_a,eq:d_a,eq:e_a,eq:theta_air_2}:
\begin{equation}\label{eq:theta_air_final}
\theta_\textrm{air}  = a_\textrm{air} + b_\textrm{air}\theta_\textrm{m,t-1}  + c_\textrm{air}\theta_\textrm{e}   +d_\textrm{air}\Phi_\textrm{sol}    + e_\textrm{air}\Phi_\textrm{HC} 
\end{equation}

Note that the notation from \citep{ISO2007} was used in this appendix. In this paper,  $T_{\textrm{air},i}^{t+1}  \leftarrow \theta_\textrm{air}$, $T_{\textrm{m},i}^{t+1} \leftarrow \theta_\textrm{m,t}$, $T_{\textrm{m},i}^{t} \leftarrow \theta_\textrm{m,t-1}$, $T_{\textrm{e}}^{t} \leftarrow \theta_\textrm{e}$, $\Phi^t \leftarrow \Phi_\textrm{sol}$, $h_{i}^{t} \leftarrow \Phi_\textrm{HC}$, such that from \cref{eq:theta_next_final,eq:kappa,eq:theta_air_final}:
\begin{equation}
\begin{bmatrix}
T_{\textrm{m},i}^{t+1}\\
T_{\textrm{air},i}^{t+1} 
\end{bmatrix}
 = \upkappa 
 \begin{bmatrix}
1,
T_{\textrm{m},i}^{t},
\textrm{T}_{\textrm{e}}^t,
\upphi^t,
h_i^t
\end{bmatrix}^\intercal
\end{equation}

This is equivalent to \cref{eq:main_heating}.

\newpage

\section{Residential energy management: commented illustrative day}\label{app:example_case_study}
Here we look in detail at the actions selected by an agent which learned to coordinate using different MARL strategies.  This is meant to illustrate how example RL $\psi$ actions translate into local energy management system behaviour. Note however that the MARL algorithms aim to generate statistically favourable outcomes when averaged over longer durations and over larger number of agents. As such, while the average outcomes are predictable, as described in \Cref{results}, this individual case is not meant to be generally representative but rather simply an example day in a stochastic environment.

\begin{figure*}[t!]
\begin{center}
\includegraphics[width=\textwidth]{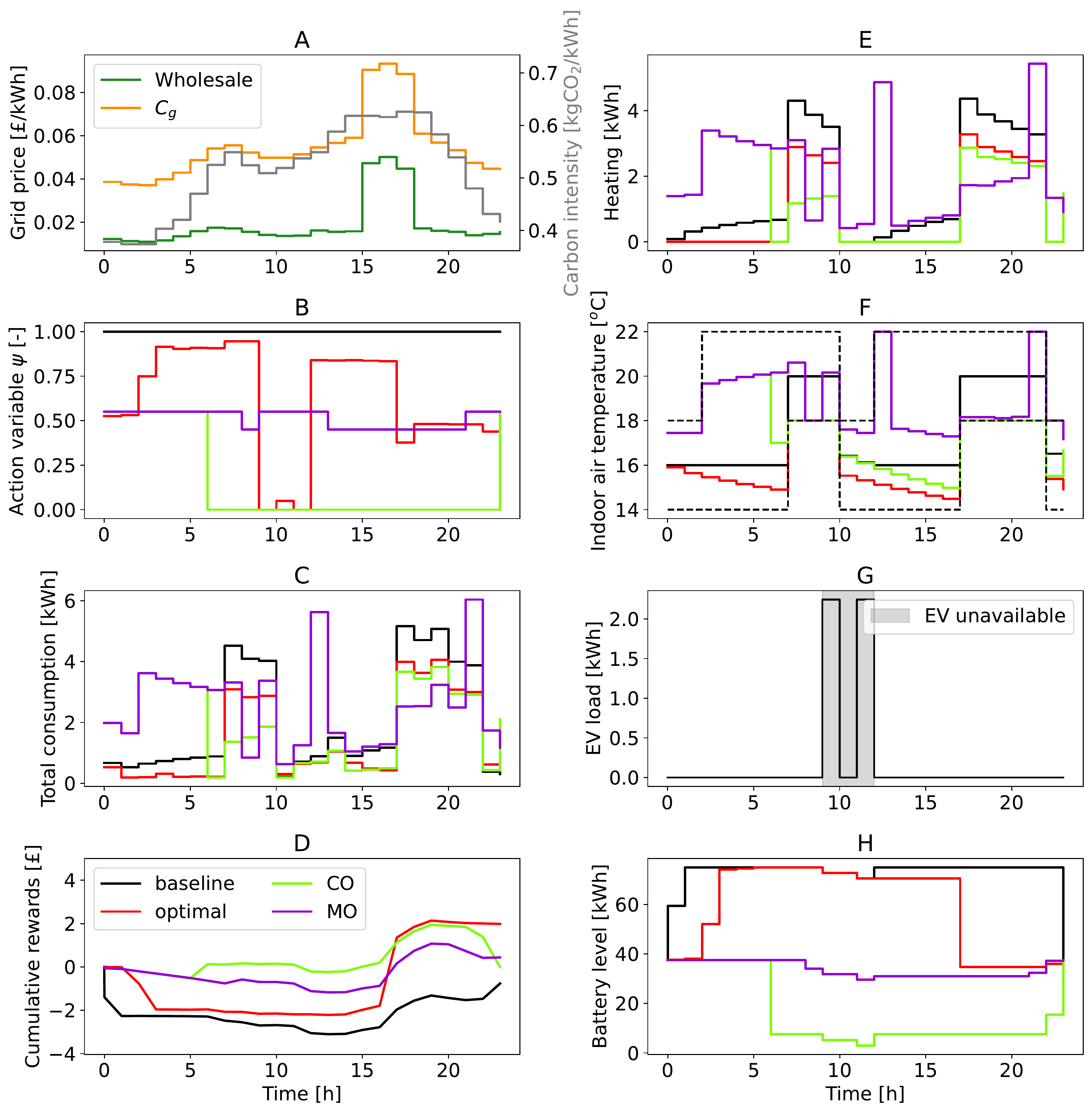}
\end{center}
\caption{Example of local home energy system variables for one agent using different policies learned ahead of implementation (optimisation-based learning using marginal rewards (MO) and optimiser state-action pairs selection counts (CO), compared with the inflexible baseline and omniscient centralised optimiser control actions}
\label{fig:example_day}
\end{figure*}

\Cref{fig:example_day} shows an example of an evaluation day during which the final policies learned is used deterministically on a day-long batch of data. Four different policies are compared: 
\begin{itemize}
\item Baseline: no flexibility used
\item Optimal: the actions selected by a central optimiser with perfect knowledge and control of all current and future variables
\item CO: a policy seeking to replicate action patterns by the optimiser by counting the number of actions taken for each grid coefficient level (state) during pre-learning (see \Cref{RLSection})
\item MO: learning from optimisations and using marginal rewards (see \Cref{RLSection})
\end{itemize}

The baseline and optimal act as reference points while the two latter policies have been identified in \Cref{results} as scalable policies when the number of agents increases. While both policies reduce costs relative to the baseline on average at scale, MO was shown to be the best-performing policy at scale. 

Subplot A shows the wholesale prices and the grid carbon intensity for the example day, as well as the resulting grid cost coefficient $C_g$ given a social cost of carbon of 70 £/tCO$_2$. This coefficient informs the choice of action $\psi$.

Subplot B shows the action $\psi$ selected by each policy over time. The CO policy seeks to imitate patterns by the optimiser -- though with more limited information than is available to the optimiser -- and takes more extreme actions, for example with maximum delaying of consumption ($\psi=0$) at time intervals with high network cost coefficient $C_g^t$, whereas the MO policy selects intermediate $\psi$ values. 

Subplot C shows the total energy consumption over time. This includes both household loads and heating consumption. MO takes intermediate $\psi$ actions and so follows more closely the baseline, non-flexible consumption profile than CO which delays more loads when taking lower $\psi$ values. Both strategies are seen to shave consumption peaks and/or displace them to lower-price time intervals. As the total household electric demand is fixed, displacing consumption does not increase total consumption. However, variation in heating loads within the acceptable temperature bounds may increase overall consumption. Thus, in this example day the MO strategy consumes more 26.1\% more energy than the baseline, though overall incurring lower costs and greenhouse gas emissions. 

Subplot D shows cumulative rewards over time each of the policies. While the CO strategy was seen to take advantage more closely of grid price differentials, overall the costs incurred are higher than with the MO strategy. With the MO strategy, savings of £1.20 are obtained compared to the baseline over the example day, corresponding to a 57.5\% reduction from baseline costs.  86.3\% of savings stem from reduced battery depreciation, 22.3\% from reduced distribution grid congestion, while grid energy costs increased by 8.6\%. In this example day the CO strategy achieved savings of £0.76, with 66.1\% stemming from reduced grid energy costs,  30.0\% from reduced battery costs, and 3.9\% from reduced grid congestion.

An interplay is thus illustrated by the two policies between the costs of battery depreciation and distribution network congestion on the one hand, and the opportunity for energy arbitrage to save on grid energy and emissions costs on the other.  Both the MO and CO strategies exhibit stable performance at scale, though converging to different types of policy. The MO policy saves more by smoothing out the charging and distribution grid utilisation profiles despite smaller savings in imports and emissions costs, while CO derives a larger advantage from the grid price differentials in grid imports, though with higher battery and distribution grid costs. The weight applied on each of those competing objectives in the objective function will have a direct impact on the policies that are learned.

Subplot E shows the heating energy profile, resulting in the temperatures in subplot F. The baseline profile maintains the median desired temperature, whereas the flexible policies can go above or below that median, within the desired temperature bounds. Both policies are more likely to absorb energy imports through heating (no marginal costs) rather than storage (battery depreciation costs) relative to the baseline policy. Consumption peaks are shaved or displaced to lower-cost time intervals with both policies.

Subplot G shows the EV at-home availability and consumption. In this example day, the electric vehicle (EV) leaves home at 9 am, consumes 2.2 kWh on an outbound trip, remains parked at its destination for 1 hour, and consumes 2.2 kWh on the inbound trip back to the home at 12 pm. The car can therefore not be charged during this time interval, and enough charge has to be available beforehand for these travelling loads. 

Subplot H shows the battery level profiles. In the baseline, the EV is charged as soon as it is plugged in, given battery capacity and charging rate constraints. The CO policy sells energy from the battery when prices increase to take advantage of the price differentials, whereas the MO policy flattens out the charging profile.

\end{appendices}

\end{document}